\documentclass[twocolumn]{aastex62}

\newcommand{\beq}{\begin{equation}\begin{aligned}}
\newcommand{\eeq}{\end{aligned}\end{equation}}

\newcommand{\msun}{M$_\odot$}

\usepackage{scalerel,amssymb}
\usepackage{amsmath}
\usepackage{enumitem}
\usepackage[]{url}
\graphicspath{{./}{figures/}}

\accepted{\today}
\submitjournal{ApJ}

\shorttitle{Radial Distribution of Satellite Systems in the LV}
\shortauthors{Carlsten et al.}

\begin{document}

\title{Radial Distributions of Dwarf Satellite Systems in the Local Volume}

\correspondingauthor{Scott G. Carlsten}
\email{scottgc@princeton.edu}

\author[0000-0002-5382-2898]{Scott G. Carlsten}
\affil{Department of Astrophysical Sciences, 4 Ivy Lane, Princeton University, Princeton, NJ 08544}

\author{Jenny E. Greene}
\affil{Department of Astrophysical Sciences, 4 Ivy Lane, Princeton University, Princeton, NJ 08544}

\author[0000-0002-8040-6785]{Annika H. G. Peter}
\affiliation{Department of Physics, The Ohio State University, 191 W. Woodruff Ave., Columbus OH 43210, USA}
\affiliation{Department of Astronomy, The Ohio State University, 140 W. 18th Ave., Columbus OH 43210, USA}
\affiliation{Center for Cosmology and AstroParticle Physics (CCAPP), The Ohio State University, Columbus, OH 43210, USA}

\author[0000-0003-4970-2874]{Johnny P. Greco}
\altaffiliation{NSF Astronomy \& Astrophysics Postdoctoral Fellow}
\affiliation{Center for Cosmology and AstroParticle Physics (CCAPP), The Ohio State University, Columbus, OH 43210, USA}

\author[0000-0002-1691-8217]{Rachael L. Beaton}
\altaffiliation{Hubble Fellow}
\affiliation{Department of Astrophysical Sciences, 4 Ivy Lane, Princeton University, Princeton, NJ 08544}
\affiliation{The Observatories of the Carnegie Institution for Science, 813 Santa Barbara St., Pasadena, CA~91101\\}

\begin{abstract}
The radial spatial distribution of low-mass satellites around a Milky Way (MW)-like host is an important benchmark for simulations of small-scale structure. The distribution is sensitive to the disruption of subhalos by the central disk and can indicate whether the disruption observed in simulations of MW analogs is artificial (i.e., numeric) or physical in origin. We consider a sample of 12 well-surveyed satellite systems of MW-like hosts in the Local Volume that are complete to $M_V<-9$ and within 150 projected kpc. We investigate the radial distribution of satellites and compare with $\Lambda$CDM cosmological simulations, including big-box cosmological simulations and high resolution zoom in simulations of MW sized halos. We find that the observed satellites are significantly more centrally concentrated than the simulated systems. Several of the observed hosts, including the MW, are $\sim2\sigma$ outliers relative to the simulated hosts in being too concentrated, while none of the observed hosts are less centrally concentrated than the simulations. This result is robust to different ways of measuring the radial concentration. We find that this discrepancy is more significant for bright, $M_V<-12$ satellites, suggestive that this is not the result of observational incompleteness. We argue that the discrepancy is possibly due to artificial disruption in the simulations, but, if so, this has important ramifications for what stellar to halo mass relation is allowed in the low-mass regime by the observed abundance of satellites. 
\end{abstract}

\section{Introduction}

One important observational benchmark with which to test models of small-scale structure formation is the radial distribution of dwarf satellites around the Milky Way (MW) and MW-like hosts. The radial distribution of luminous, low-mass satellites is sensitive to the physics of reionization \citep[e.g.][]{kravtsov2004,dooley2017} and to the disruption of the subhalos that host the satellites by the central primary \citep[e.g.][]{donghia2010,gk_lumpy,kelley2019,samuel2020}. Of particular importance is understanding whether the disruption of subhalos is physical or an artificial feature of the simulations \citep[due to e.g. low resolution effects;][]{vdb2018a, vdb2018b}. Tidal stripping and disruption of subhalos are both integral parts of the baryonic solutions to the well-known ``small-scale challenges'' to $\Lambda$CDM \citep[e.g.][]{gk_lumpy, gk2019}, and, hence, it is critical to understand them fully. 

Previous comparisons between observations of the MW satellites and the predictions from $\Lambda$CDM simulations have produced somewhat baffling results. Comparisons with the classical satellites ($M_*\gtrsim10^5$~\msun) indicate that the MW satellites are significantly more radially concentrated than the most massive dark matter (DM) subhalos in dark-matter-only (DMO) simulations \citep{kravtsov2004, lux2010, yniguez2014}. Many studies have argued that reionization can help in this regard as the distribution of \textit{luminous} subhalos will be more centrally concentrated than the overall population of subhalos \citep[e.g.][]{moore2001,taylor2004, kravtsov2004, font2011, starkenburg2013, barber2014, dooley2017}. The earliest forming subhalos will be the ones most likely to be luminous and are more concentrated near the host. While the halo mass scale at which reionization starts to significantly suppress galaxy formation is still uncertain, recent simulations indicate that it is at (or well below) the low-mass end of halos expected to host classical-sized satellites \citep{sawala2016b, wheeler2019}. Thus, reionization does not appear to be a viable explanation for the concentration of the MW classical satellites. The fully hydrodynamic simulations presented in \citet{samuel2020} showed a similar result as \citet{yniguez2014} with the MW still substantially more concentrated than simulations in several metrics. Unfortunately, the results in comparison to the classical satellites are fundamentally limited by the low statistics offered by the $\sim10$ MW classical satellites. 

One way to increase statistics is to consider the ultra-faint dwarf (UFD) satellites of the MW as well, of which there are now $\sim50$ known. The MW UFDs seem to also be more concentrated than DMO simulations would predict, particularly when disruption by the baryonic disk of the host is included. 
Investigating the simulations of \citet{gk_lumpy}, \citet{kim2018} found that there were not enough subhalos that survived the enhanced disruption by the disk to host the known MW UFDs at small radii, assuming a reasonable stellar halo mass relation (SHMR).
A similar conclusion was reached by \citet{graus2019}. \citet{graus2019} noted that the close-in MW UFDs could be explained if \emph{very} low mass subhalos were populated. These low mass subhalos are well below the usual cutoff for luminous satellites due to reionization suppression of galaxy formation \citep[e.g.][]{bullock2000, somerville2002, okamoto2008, okamoto2009}. It is still unclear how galaxies could form in these halos, but if they do, there is no reason to suspect they would be centrally concentrated. Thus, there should be a very large number of UFDs in the outskirts of the MW virial volume awaiting discovery \citep[see also][for estimates]{kim2018}. 

Using the UFDs to test simulation predictions with observations comes with its own problems, however. The observational census of UFDs is radially incomplete due to their intrinsic faintness and small size \citep[e.g.][]{koposov2008, walsh2009} and this is biasing the observed radial distribution. Also, as the UFDs likely reside in very low-mass subhalos, the resolution of the simulations becomes a major concern. In the end, the MW is still only one system, and it is difficult to draw broad conclusions on the radial distribution of satellites from the MW alone. 

In this paper, we take a complementary approach to the work presented above by studying the observed radial distributions of classical-mass satellites around many hosts in the Local Volume (LV). By comparing multiple MW-like hosts together, we are able to get far better statistics than with the MW alone. The basic question that this paper tries to answer is ``how well do the spatial distributions of dwarf satellite systems created in modern simulations match those observed for the MW and MW-analogs in the LV?''. We compare the observed systems to a wide range of recent simulations, including big-box cosmological hydrodynamic simulations with many simulated MW-like hosts, high-resolution DMO zoom simulations (both with and without an included disk potential), and high-resolution hydrodynamic zoom simulations that include only a handful of MW-like hosts. Using these very different simulations allows us to explore how much the simulation results depend on the properties of the simulation, including resolution. In this paper, we thus perform the first comparison between a \emph{population} of observed satellite systems and a \emph{population} of simulated analogs. This has only recently been made possible with the creation of all of these simulation suites and the observations required to characterize the satellite systems of nearby MW analogs. By studying the radial distributions of a population of satellite systems, it might also be possible to learn about the scatter between satellite systems \citep[i.e. why is the radial distribution of M31 so different from that of the MW?][]{yniguez2014, willman2004}.

This paper is structured as follows: in \S\ref{sec:data} we present the observational data sets of LV satellite systems, in \S\ref{sec:models} we list the different simulation suites that we compare with, and in \S\ref{sec:comparison} we show the comparison of observations with models. In \S\ref{sec:disc}, we discuss possible causes for the discrepancy that we find, including possible caveats related to both the observations and simulations, and we conclude and outline directions for future work in \S\ref{sec:concl}.

\section{Sample of Satellite Systems}
\label{sec:data}
In this work, we compare simulations with a sample of 12 well-characterized observed satellite systems around massive hosts in the Local Volume ($D\lesssim10$ Mpc). Characterizing these satellite systems has been the result of the combined effort of multiple groups over the last several years. Still, due to the inherent faintness of dwarf satellites and the large areas that need to be surveyed, only a small number of massive hosts have been surveyed to the point that the inventory of `classical'-sized satellites is likely complete for a large fraction of the host's virial volume. Most of the difficulty is in measuring the distances to these low-mass galaxies to confirm that they are actually physically associated to a specific host. Several previous studies \citep[e.g.][]{sbf_m101, bennet2019, lv_lfs} have shown that the contamination from background interlopers along the line of sight can be significant ($>80$\%).  

For this study, we use the compilation of LV satellite systems given in \citet{lv_lfs}. That work uses the catalog of satellite candidates around ten LV massive hosts from \citet{LV_cat} and confirms satellites using distances measured via surface brightness fluctuations \citep[SBF;][]{sbf_calib}. Six of the hosts we consider here were characterized in this work (NGC 1023, NGC 4258, NGC 4631, M51, M104, and NGC 4565)\footnote{The other four hosts considered in that work had significantly less survey area coverage and are not considered here.}. The first five of these hosts are mostly complete down to a satellite luminosity of $M_V\sim-9$ and surface brightness of $\mu_{0,V}\sim26.5$ mag arcsec$^{-2}$ within the inner $\sim150$ projected kpc \citep[see][for more detailed quantification of completeness]{LV_cat}. For these hosts, there were still a few candidates where the SBF results were ambiguous. We treat these candidates as `possible' satellites and always give a spread of possible results (both including them as satellites and not). The sixth host, NGC 4565, had good area coverage and deep survey data, but due to its larger distance ($D=11.9$ Mpc) and worse seeing, the SBF distance results were ambiguous for all but the brightest candidates. All of the candidates brighter than $M_V=-12$ were confirmed by SBF or redshift such that there were no ambiguous candidates in NGC 4565 at this luminosity. Thus, we include NGC 4565 as well, but we note that its satellite system is only characterized down to $M_V=-12$. 

For NGC 4631, we have used DECaLS \citep{decals} with Gemini$+$GMOS followup to identify more satellites outside of the original footprint of \citet{LV_cat}. Its satellite system is now likely complete to $\sim200$ kpc. We describe the extra data in Appendix \ref{app:extra_data}. 

In addition to the six hosts from \citet{lv_lfs}, we include the six other nearby hosts that have had their satellites well surveyed in previous work. These six are the MW, M31, Centaurus A (CenA; NGC 5128), M81, M94, and M101. The specific lists of satellites that we consider in each of these systems can be found in the appendix of \citet{lv_lfs}. The MW satellite list comes from \citet{mcconnachie2012} and uses distances from \citet{fritz2018}. We do include the dSph Sgr, but note that because it is currently disrupting objects like it may no longer be identified in simulations by halo finders. It is not clear what the correspondence is between simulations and observations for various stages of tidal disruption, and whether systems like this can be meaningfully compared with simulations. Overall, our results do not change qualitatively with the inclusion of Sgr, and we discuss this more below.  The M31 satellites come from \citet{mcconnachie2018} with distances primarily from \citet{weisz2019}. The CenA satellites come from the compilations of \citet{muller2019} and \citet{crnojevic2019}. The satellites of M81 are taken from \citet{chiboucas2013}. M94 satellites come from \citet{smercina2018}. Finally, the M101 satellite system comes from the work of \citet{bennet2017}, \citet{danieli101}, \citet{sbf_m101}, and \citet{bennet2019}. These works surveyed the satellites out to $\sim$200 kpc. We have used DECaLS \citep{decals} imaging to identify 2 more satellites of M101, which we have confirmed with SBF, completing its satellite system out to 300 kpc. We describe this in Appendix \ref{app:extra_data}.

A detailed discussion of the completeness of each of these satellite systems can be found in \citet{lv_lfs} and references therein. In brief, we assume that the MW and M31 systems are complete to classical satellites ($M_V\lesssim-8$) within the inner 300 kpc. M101 is complete to $M_V\sim-8.5$ within the inner 300 kpc. M81 is likely complete to better than $M_V\lesssim-9$ within the inner projected 250 kpc. CenA is likely complete within the inner projected 200 kpc to about $M_V\lesssim-9$. M94 is complete within only the inner 150 kpc at this luminosity limit. For the six hosts from \citet{lv_lfs}, we assume the satellite systems are complete to $M_V\sim-9$ (with the exception of NGC 4565 noted above), and we use the actual survey footprints [see Fig 1 of \citet{LV_cat}] to characterize the areal completeness. 

As in \citet{lv_lfs}, we only consider satellites that have surface brightness above $\mu_{0,V}\sim26.5$ mag arcsec$^{-2}$. Significantly lower surface brightness satellites (even with total luminosity of $M_V\sim-9$) are detectable from resolved stars around the MW and M31, but these would not be detectable in the majority of the LV hosts we consider here.    Additionally, like Sgr, several of the extremely LSB satellites (e.g. AndXIX, \citet{mlmc2020}) are clearly undergoing tidal stripping, and it is unclear if these satellites are appropriate to include in the comparison with simulations below. The subhalos hosting such stripped systems might not be recognized by the halo finders used in the simulations, as discussed more below. Since Sgr is much more massive (2-3 orders of magnitude), it is more likely than these objects to be recognized as a subhalo.

Properties for all 12 hosts considered in this work can be found in \citet{lv_lfs} and \citet{LV_cat}. These are all massive hosts with stellar mass ranging from roughly 1/2 that of the MW to $\sim5\times$ that of the MW. As discussed in \citet{lv_lfs}, the hosts naturally split in two groups based on halo mass. The categorization of each host into these two groups was based on stellar mass, circular rotation speed, and any available estimate of the halo mass from satellite dynamics. The low-mass group are all very similar to the MW, and we will refer to these as `MW-like' or `MW-analogs'. These include the MW, M31, M94, M101, NGC 4631, NGC 4258, NGC 4565, and M51. We estimate the halo masses of these hosts are in the range $\sim0.8-3\times10^{12}$~\msun. We refer to the more massive hosts as `small group' hosts and include M81, CenA, NGC 1023, and M104. These hosts correspond to halo masses in the range $\sim3-8\times10^{12}$~\msun. It is important to compare the observed hosts with simulated hosts of similar mass; we often consider each group of observed hosts separately and compare each individually with the appropriate simulated hosts.

This sample of 12 hosts is representative of the overall sample of massive hosts in the Local Volume. It includes nine star-forming spiral galaxies, two ellipticals, and one lenticular. We note that this sample is almost \emph{volume limited} for hosts within $D\sim8.5$ Mpc. Our study contains 8 out of 11 massive ($M_\star\gtrsim1/2\times M_\star^{\mathrm{MW}}$) hosts with $|b|>15^\circ$ within this volume. The missing hosts are NGC 5236 (M83), NGC 253, and NGC 4826 (M64).

\section{Models}
\label{sec:models}
We compare the observations with a wide variety of simulations both to improve the statistics and to see how the inferences we draw regarding these systems depend on the specifics of the simulation. We include both big-box cosmological simulations that contain many MW-like hosts but at a low resolution, and zoom-in cosmological simulations that focus on only a single MW-sized host but at a much higher resolution. We include DMO simulations, DMO simulations with an added disk potential, and fully hydrodynamic simulations. This allows us to explore the effect of subhalo disruption by the central disk on the spatial distribution of satellites.

For all of the simulation suites except for the zoom hydrodynamic simulations, we only use the dark matter (DM) halo catalogs from the simulations. To populate these subhalos with luminous galaxies we could use a stellar halo mass relation (SHMR) to assign a stellar mass to each subhalo. \citet{lv_lfs} found decent agreement between the observed satellite LFs in the current sample and the simulated LFs assuming the SHMR of \citet{gk_2017}, a commonly adopted SHMR from the literature. However, there is still significant uncertainty in what the true SHMR is for this mass range. Therefore, in order to keep the results regarding the radial distribution as general as possible, we do not use a SHMR to populate subhalos. Instead, we select the $n$ most massive subhalos that fall in the survey footprints where $n$ is the number of observed satellites (either in a particular observed host or averaged over several observed hosts). More specific details are given with each comparison. When considering subhalos masses, we always consider the peak virial mass over the subhalo's history, $M_\mathrm{peak}$.

For the cosmological simulation, we use the public IllustrisTNG-100\footnote{\url{https://www.tng-project.org/data/}} simulation \citep{tng1, tng2, tng3, tng4, tng5, tng6}. We use the DM halo catalogs from the full hydrodynamic simulation run and not the DMO simulation run. The full hydro simulation should include the effect of subhalo disruption by the baryonic disk of the host. The baryonic mass resolution of TNG is $\sim10^6$ \msun, which means that the satellites of the luminosity we are interested in ($M_V<-9$ mag) are not resolved. However, the DM particle mass of $7.5\times10^6$ \msun\ means that subhalos hosting the satellites we are focusing on ($\sim5\times10^9$ \msun) will contain $\sim1000$ particles at infall. From the $100^3$ Mpc$^3$ box, we select host halos as described in \citet{lv_lfs}. In brief, when we compare the simulations individually with each observed host, we select the simulations to have stellar mass consistent with the observed host, using the central's stellar mass predicted in the hydrodynamic results. Each TNG host is given a probability to be included given by a Gaussian distribution in log stellar mass that is centered on the stellar mass of the observed host with spread 0.1 dex. We assume 0.1 dex is an appropriate estimate of the error in determining the stellar mass of nearby massive galaxies. The resulting distribution of stellar masses of the selected TNG hosts is peaked at the observed stellar mass of the LV host but allows some spread due to measurement uncertainty. When comparing with all observed hosts together, we select TNG hosts based on halo mass in the range $0.8\times10^{12}<M_{200}<8\times10^{12}$ \msun. 

For the zoom-in simulations, we use three separate simulation suites. These offer significantly higher resolution than IllustrisTNG, which allows us to explore the effect of resolution on the model predictions. The first is the ELVIS\footnote{\url{http://localgroup.ps.uci.edu/elvis/index.html}} \citep{elvis} suite of DMO simulations. This suite includes 24 isolated MW-sized hosts and a further 24 that are in a paired Local Group (LG) configuration. We consider all 48 of these hosts in the same way. With a DM mass resolution of $1.6\times10^5$~\msun, this simulation is significantly higher resolution than IllustrisTNG. The second zoom simulation we include is the PhatELVIS\footnote{\url{http://localgroup.ps.uci.edu/phat-elvis/}} suite \citep{kelley2019} of DMO simulations. These simulations are distinct from the ELVIS suite both by being even higher resolution (DM mass resolution of $3\times10^4$~\msun) and also that they account for the enhanced disruption of subhalos due to the baryonic disk of the host. A gravitational potential grown to match that of the MW's disk is artificially put into the simulations. PhatELVIS includes 12 simulated hosts. Both the ELVIS and PhatELVIS suites are DMO, and so we just use the DM subalo catalogs, as described above. The final zoom simulation we include is the full hydrodynamic simulations from the NIHAO project\footnote{\url{http://www2.mpia-hd.mpg.de/~buck/\#sim_data}} \citep{buck2019}. There are only 6 hosts in this suite, but the simulations are very high resolution with DM particle mass of $\sim10^5$ \msun\ and star particle mass of $\sim10^4$ \msun. Since the simulations are hydrodynamic we take the luminous galaxy catalog directly from the simulations. The lowest-mass satellite we consider in this paper has $M_*>10^5$~\msun, which will be at least marginally resolved in the NIHAO results. This is a similar resolution used in the FIRE-2 simulations recently presented by \citet{samuel2020}. All of the zoom-in simulations are of hosts roughly the halo mass of the MW, so we only compare these simulations to the MW-like observed hosts. Specific details (including masses) of each of the simulated zoom-in hosts can be found in the respective publications.

\section{Results}
\label{sec:comparison}
In this section, we compare the observed satellite systems to the simulated systems. Throughout this paper, we focus primarily on the shape of the radial distribution of satellites and not the absolute radial distribution. We compare the observations with the models in many different ways, using distinct metrics, to try to get a complete understanding of how well they agree. Due to the large number of simulated hosts, we primarily compare the observations to the IllustrisTNG results, and secondarily consider the other simulation suites to show consistency. We start by considering the normalized 2D (projected) radial distributions of all of the observed hosts compared to the simulations. We then compare the normalized 3D distribution of satellites around the MW and M31 (the only observed hosts where we have 3D information on the satellite positions). Finally, we explore different ways of parameterizing the shape of the radial distribution.

\begin{figure*}
\includegraphics[width=\textwidth]{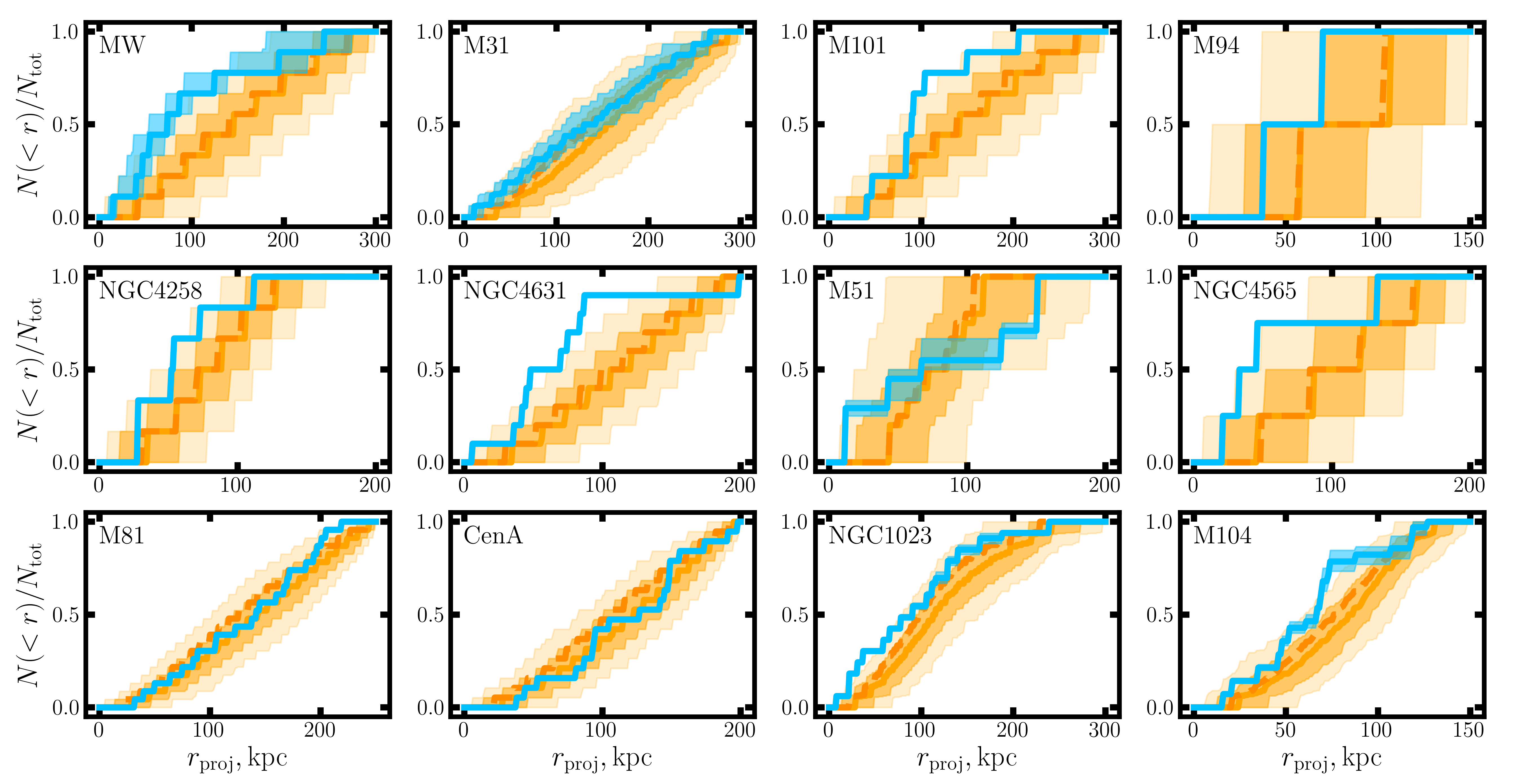}
\caption{The 2D (projected) cumulative radial distributions of satellites ($M_V<-9$) around the 12 LV hosts considered in this work. All profiles are normalized by the total number of satellites in the surveyed area. Observed systems are shown in blue while the analogous simulated systems from IllustrisTNG are shown in orange. The dark-orange dashed line shows the median profile for the ELVIS DMO simulations. The simulations are forward-modelled using the area completeness of the surveys for each host. For the simulations, the $n$ most massive subhalos falling in the survey region are identified as the luminous satellites, where $n$ is the observed number of satellites in the survey region for that specific host.  For the MW and M31, the shaded region shows the effect of different projection angles on the radial profile. For the other LV hosts, the shaded region encompasses any uncertainty in membership of candidate satellites without distance information. The bottom row of hosts are the more massive `small-group' hosts.} Note that the range of the $x$-axis is different for different hosts, depending on the radial coverage of its satellite census.  
\label{fig:indiv}
\end{figure*}

\begin{figure*}
\includegraphics[width=\textwidth]{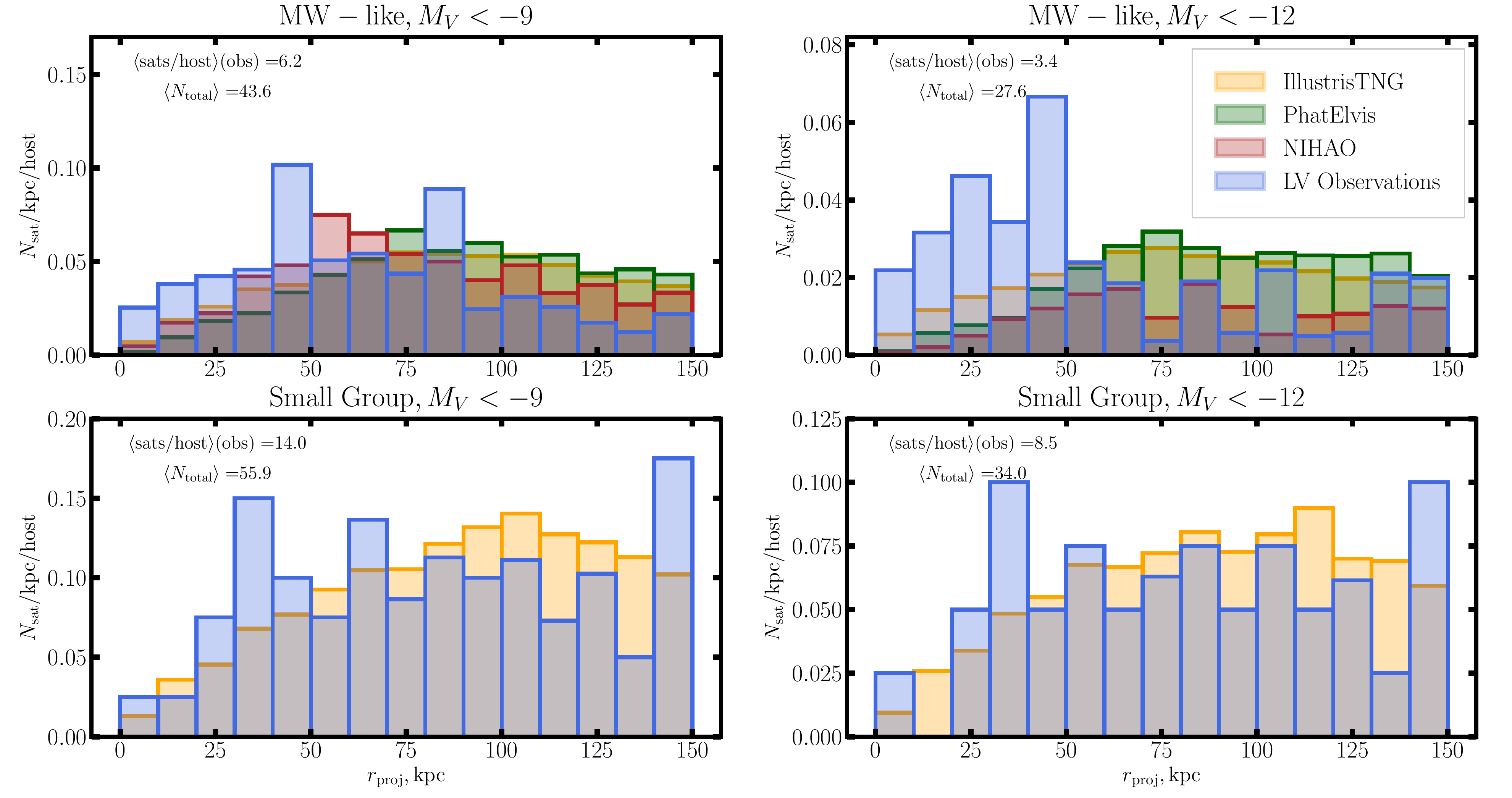}
\caption{Histogram of the projected satellite separation ($r_{\mathrm{proj}}$) for all observed hosts together compared against the simulated hosts. The observation histograms are normalized such that the total area is the average number of satellites per host. These numbers are given in the top left corner of each panel, along with the average total number of observed satellites contributing to each histogram (which is not necessarily an integer due to the effects of different projecting angles for the MW and M31). The top panels are for the MW-like hosts while the bottom are for the more massive `small-group' hosts. The left panels are for all satellites $M_V<-9$ while for the right panels, only $M_V<-12$ satellites are included. We restrict to satellites within the inner 150 kpc projected, which is the extent of the radial coverage for most of the hosts.}
\label{fig:rproj}
\end{figure*}

\subsection{2D Projected Radial Distributions}
\label{sec:2d}
In Figure \ref{fig:indiv}, we show the radial profiles for each of the 12 observed hosts that we consider in this work. We only consider satellites brighter than $M_V<-9$ and assume that all hosts are complete in luminosity down to this level (except NGC 4565 for which we use a luminosity limit of $M_V<-12$). Each host is compared with the simulated systems from IllustrisTNG. The IllustrisTNG hosts are selected based on stellar mass to be consistent with each observed host, as described in \S\ref{sec:models}. We also show the radial profiles predicted by the DMO ELVIS simulation suite. The simulated systems are mock-observed at the distance of each host, and the observational area selection function for each host is used to select which simulated subhalos would be observed. We select subhalos whose line-of-sight (LOS) distance from the observer is within 500 kpc of the host. This accounts for the fact that for most of the hosts, the distances available for the satellites are not high enough precision to probe the 3D structure of the group. SBF distances are accurate to $\sim15$\% \citep[e.g.][]{sbf_calib} while \emph{HST} TRGB distances are accurate to $\sim5$\% \citep[e.g.][]{danieli101, bennet2019}. At a host distance of 7 Mpc, these correspond to uncertainties of $\sim1000$ and $300$ kpc, respectively. It is entirely possible that some of the `confirmed' satellites of these hosts are actually near-field galaxies that project onto the host but are outside of the virial volume of the host. Thus, we account for the line-of-sight (LOS) uncertainty by including subhalos within 500 kpc LOS of the host. For the MW and M31, we have detailed 3D positions for the satellites, and we use that to mock `re-observe' these systems at a distance of 7 Mpc. This allows us to explore the effect of the observing angle on the radial profile. For the hosts from \citet{lv_lfs} that have some unconfirmed candidate satellites, the uncertainty in membership is accounted for as a spread in possible radial profiles. Specifically, each possible combination of the unconfirmed members is considered, and that many radial profiles are generated. We plot the median and $\pm1\sigma$ spread in these profiles. 
 
Since the scatter in the profile between hosts will depend on how many satellites each host has, for a fair comparison, we select the same number of subhalos from each simulated host as is observed for a particular observed host. The $n$ most massive subhalos (considering peak mass) that fall in the survey footprint are selected as the luminous satellites where $n$ is the number of observed satellites for a specific host.

While there clearly is significant scatter between the observed hosts in Figure \ref{fig:indiv}, the observed hosts appear to be generally more concentrated than the simulated hosts. Several of the observed hosts (e.g. the MW, M101, NGC 4258, NGC 4631, NGC 4565, and NGC 1023) have their profiles at or just within the $-2\sigma$ (i.e. more centrally concentrated) scatter in the simulations whereas no host is correspondingly outside the $+2\sigma$ (i.e. less concentrated) scatter in the simulations. The `small-group' hosts are less discrepant with the simulations. Indeed, both M81 and CenA closely follow the median simulated profile.

Another way to assess the concentration of the satellite population is the histogram of the satellites' projected separations, $r_\mathrm{proj}$ from their hosts. In Figure \ref{fig:rproj}, we show the distribution of all satellite projected separations across all hosts combined. The histograms are normalized such that the total area under the curve is the average number of satellites per host\footnote{The bins are simply chosen to be 10 kpc wide. While different binning schemes could significantly change the appearance of the distributions, we discuss the significance of the disagreement of the distributions using metrics that do not require binning below.}. Only satellites within $r_\mathrm{proj}<150$ kpc are included. We consider the MW-like and small-group hosts separately and look at all ($M_V<-9$) satellites and just the brighter ($M_V<-12$) satellites. This luminosity threshold is motivated by the completeness limit of the SAGA Survey \citep{geha2017} and enables comparison with this survey, because it is the closest in spirit to this one. The histograms of the observations are averaged over the viewing angle for the MW and M31 and also averaged over the uncertainty in membership for some candidates. 

The IllustrisTNG hosts are selected based on halo mass. Halos in the mass range $0.8\times10^{12}<M_{200}<3\times10^{12}$ \msun\ are compared with the `MW-like' observed hosts, and halos in the range $3\times10^{12}<M_{200}<8\times10^{12}$ \msun\ are compared to the `small-group' hosts. The simulations are forward-modelled to include the effect of the survey footprints of the observed hosts in a similar way as in Figure \ref{fig:indiv}. For each simulated host, one of the observed hosts is selected at random, and that simulated host is forward modelled through the area selection function of that observed host. The decrease in satellites at large $r_\mathrm{proj}$ in the simulated hosts is largely due to some of the observed hosts not being surveyed fully to 150 kpc. For each NIHAO and PhatELVIS host, 50 different viewing angles are taken. For the IllustrisTNG and PhatELVIS simulations, the $n$ most massive subhalos in the survey footprint are selected as the satellites where $n$ is the average number of observed satellites above the luminosity limit. The average number of observed satellites is given in the upper left corner of each histogram in Figure \ref{fig:rproj}, along with the average total number of satellites amongst the hosts shown in each panel. The total number of satellites is not necessarily an integer due to the effects of different projecting angles for the MW and M31 and uncertain membership for some of the other hosts.  We use the NIHAO hydro results to take satellites above the same luminosity threshold as used for the observations.

The observed MW-like hosts have noticeably shifted distributions of $r_\mathrm{proj}$ compared to the simulations. Restricting to only the bright ($M_V<-12$) satellites makes this discrepancy \textit{significantly} more noticeable. On the other hand, the more massive `small-group' hosts have observed $r_\mathrm{proj}$ distributions that are only slightly flatter than the simulated hosts.

To assess the significance of the discrepancy with the MW-like hosts, we use a two-sample KS test between the observations and simulations. To account for the different viewing angles of the MW and M31 and uncertain membership, we consider many different realizations of the observed hosts (with different viewing angles and different memberships) and calculate the KS statistic with the ensemble of PhatELVIS simulations. We find median $p$-values that the two samples are drawn from the same distribution of $5\times10^{-3}$ for all ($M_V<-9$) satellites and $5\times10^{-4}$ for the bright ($M_V<-12$) satellites, when we just consider one viewing angle per PhatELVIS host. When compared to the TNG simulations, we find similar median $p$-values of $4\times10^{-3}$ and $5\times10^{-4}$, respectively. The PhatELVIS and TNG simulations do show fair agreement with each other with $p$-values that their radial distributions are drawn from the same underlying distribution of $\sim0.8$ (using one viewing angle per PhatELVIS host).

\subsection{3D Radial Distributions}
\label{sec:3d}

\begin{figure*}
\includegraphics[width=\textwidth]{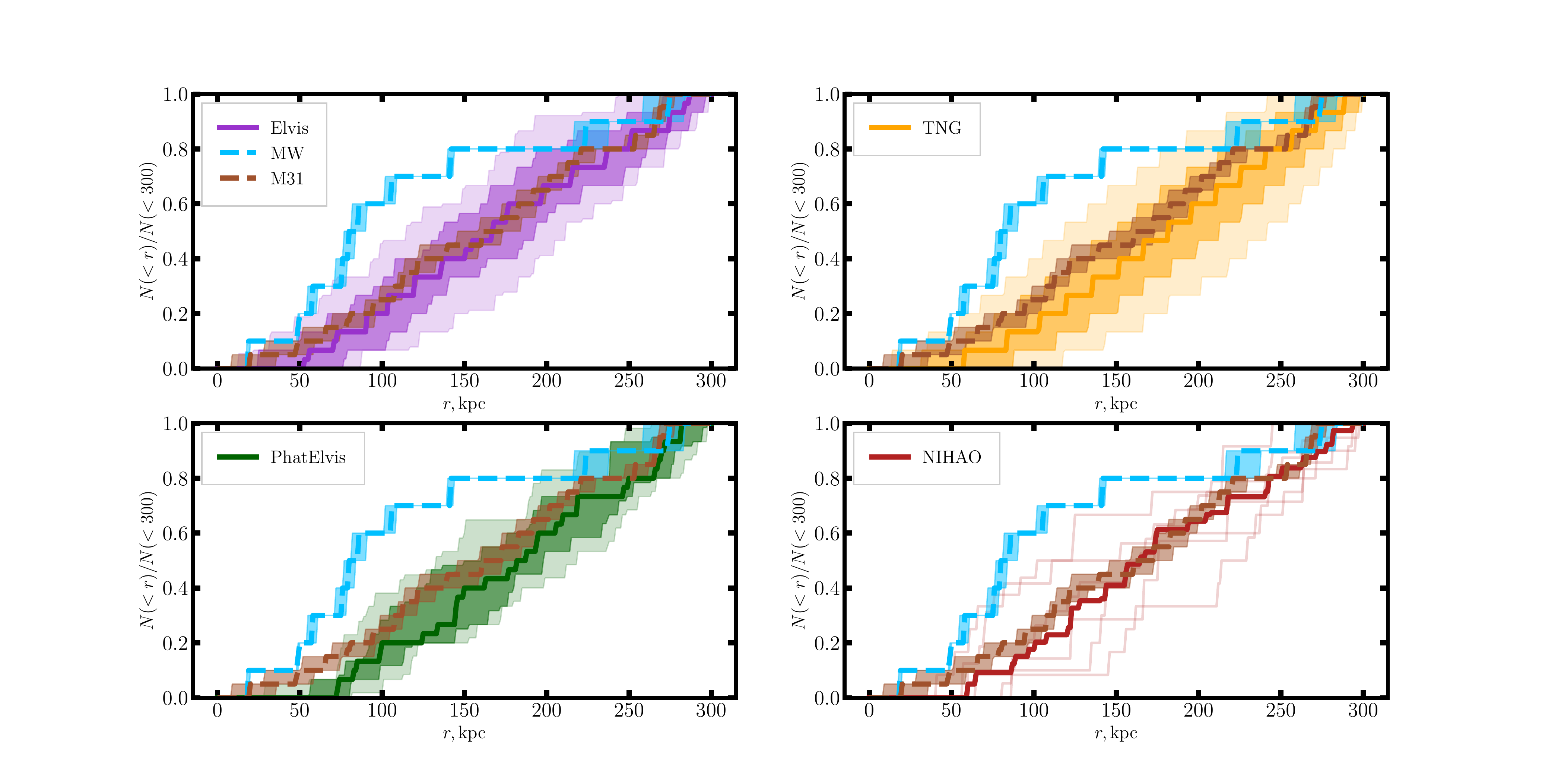}
\caption{The 3D radial distribution of MW and M31 satellites ($M_V<-8$) compared with four different simulation suites. For the DMO simulations, the 15 most massive subhalos are selected as satellites. The profiles are cumulative and normalized by the number of satellites within 300 kpc. The shaded bands for M31 and the MW account for distance uncertainties in the satellites. The shaded bands for the simulations denote the $\pm1\sigma$ and $\pm2\sigma$ spread in the simulation results. As there are only 6 hosts in the NIHAO suite, they are plotted individually, and the thick line shows the median.}
\label{fig:r3d}
\end{figure*}

Figure \ref{fig:r3d} shows the 3D radial distributions of the classical ($M_V<-8$)\footnote{Note, we still implement the $\mu_{V,0}<26.5$ mag arcsec$^{-2}$ cut for consistency. The result, however, is unchanged if we include satellites of all surface brightness.} satellites around M31 and the MW compared to the four simulation suites that we consider in this paper. The distributions are cumulative and normalized to the number of satellites within $r<300$ kpc.  For the DMO simulations, the 15 most massive (peak mass) subhalos are selected for each host, roughly in between the number of classical satellites around the MW and that around M31. The TNG hosts are selected based on halo mass in the range $0.8\times10^{12}<M_{200}<3\times10^{12}$ \msun. For the NIHAO hosts, the hydrodynamic results are used, and satellites with $M_V<-8$ are selected. 

There are two interesting things to note from this plot. The first is that the different simulation results look remarkably similar to each other. Ostensibly this means that at the satellite mass we focus on, the resolution of the simulations is not affecting the results (or is affecting them all in the same way), and that even the low-resolution IllustrisTNG results are converged at this mass scale. We consider this point in more detail in \S\ref{sec:disc}. The biggest difference between the simulations is that the PhatELVIS simulated hosts have significantly fewer satellites within 100 kpc than the other simulation suites. This is due to subhalo disruption by the disk potential that \citet{kelley2019} injected into the simulations; ELVIS, however, has no added disk. Both the Illustris and NIHAO simulations should have this effect because the host will form a disk in these hydrodynamic simulations, so it is unclear why the PhatELVIS simulations show a much more pronounced deficit of subhalos in the inner regions. 

The second thing to note from Figure \ref{fig:r3d} is that the MW's satellite distribution is significantly more centrally concentrated than any of the simulation results. It is well outside of the $2\sigma$ regions for all of the simulations. This confirms the result of \citet{yniguez2014}, who used the ELVIS simulations. Most hosts have significant populations of satellites outside of $r=150$ kpc whereas the MW has only two in this luminosity and surface brightness range. M31, on the other hand, appears to have a radial distribution fully consistent with the simulations. Both \citet{yniguez2014} and \citet{samuel2020} argue that this unusual concentration indicates that undiscovered MW satellites exist far out in the virial volume, awaiting discovery. We address the point of completeness of the MW satellite census in \S\ref{sec:disc_obs} and argue that this is unlikely and, even if true, unlikely to actually help ease the discrepancy. 

The fact that the MW is unusually centrally concentrated in both 2D (projected) and 3D is suggestive that the other overly concentrated LV hosts would be similarly concentrated in 3D. Interestingly, we note that the MW appears to be a \textit{significantly} more extreme outlier in 3D than 2D which is also suggestive of what the other LV hosts might look like in 3D.

\begin{figure*}
\includegraphics[width=\textwidth]{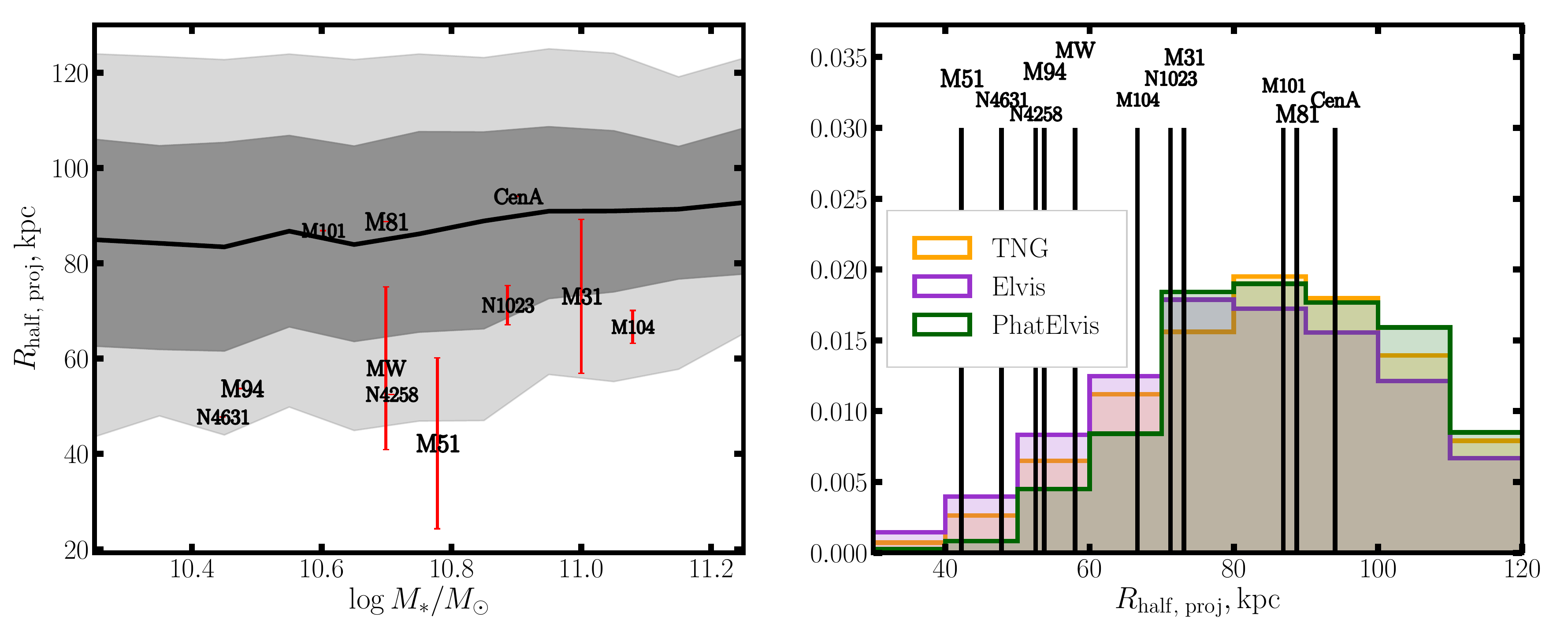}
\caption{The radius encompassing half of the satellites for 11 observed hosts versus the simulated hosts. Only satellites brighter than $M_V<-9$ and within a projected 150 kpc of their host are considered. On the left, the median satellite radius is plotted against the stellar mass of the host. The contours in the background show the $\pm1,2\sigma$ results for the IllustrisTNG hosts. The errorbars for M31 and the MW show the effect of different observing angles while the errorbars on the other hosts indicate the effect of uncertain membership for the subset of candidate satellites without distance measurements. On the right, the median satellite separation for the hosts are compared against the simulations in histogram form. The results from the ELVIS and PhatELVIS suites are also shown.}
\label{fig:r50}
\end{figure*}

\subsection{Satellite Concentration and Host Stellar Mass}

To explore a different metric describing the shape of the radial profiles, in Figure \ref{fig:r50} we show the radius that contains half of the satellites for the observed and simulated hosts. This $R_{\mathrm{half}}$ is calculated, essentially, as the median satellite projected separation from the host. To compare all observed hosts together, we only consider satellites within a projected 150 kpc from their host. Only satellites brighter than $M_V<-9$ are considered and, thus, we do not include NGC 4565 in this assessment. On the left, the half-satellite radius is plotted against the stellar mass of the host. The observed hosts (points) are compared with the simulated IllustrisTNG hosts in the background. The IllustrisTNG hosts are selected by halo mass in the range $0.8\times10^{12}<M_{200}<8\times10^{12}$ \msun. The average observed number of satellites within 150 kpc and with $M_V<-9$ (including the projection effects of the MW and M31 and the effect of uncertain membership) is 6 per MW-like host and 14 per small-group host. Thus, we select the 6 most massive subhalos for each TNG MW-like host ($0.8\times10^{12}<M_{200}<3\times10^{12}$ \msun) and the 14 most massive for each small-group host ($3\times10^{12}<M_{200}<8\times10^{12}$ \msun) to compare with the observed satellites. For the IllustrisTNG hosts, we use the stellar mass of the host reported by the hydrodynamic simulation results.  The IllustrisTNG hosts are all mock-observed at $D=7$ Mpc (roughly the average distance of the observed hosts). For each simulated host, one of the observed hosts is chosen at random and that host's survey area selection function is applied to the simulated host. On the right, the half-satellite radii of the observed hosts are compared in histogram form against the IllustrisTNG, ELVIS, and PhatELVIS simulated hosts. For the ELVIS and PhatELVIS hosts, the 6 most massive subhalos in the survey footprints are selected, and 100 viewing angles are taken for each host. 

From Figure \ref{fig:r50}, we see that for satellites within a projected 150 kpc of their host, the median satellite separation is $\sim90$ kpc for the simulations but is closer to $\sim50-70$ kpc for the observed satellite systems. The observed hosts are systematically more centrally concentrated than the simulated hosts at the $1-2\sigma$ level. The higher-mass observed hosts have their satellites at somewhat larger radii, in agreement with Figure \ref{fig:indiv}. The three different simulation suites that we consider all show similar satellite spatial distributions. The ELVIS hosts are slightly more centrally concentrated than the IllustrisTNG or PhatELVIS hosts, but that is easily understood due to lack of any central disk. A few of the most massive hosts (M81, CenA, and M31) have spatial distributions characteristic of the simulated hosts, but the majority of the observed hosts are more concentrated, and, most importantly, \textit{none} are less concentrated.

\begin{figure}
\includegraphics[width=0.47\textwidth]{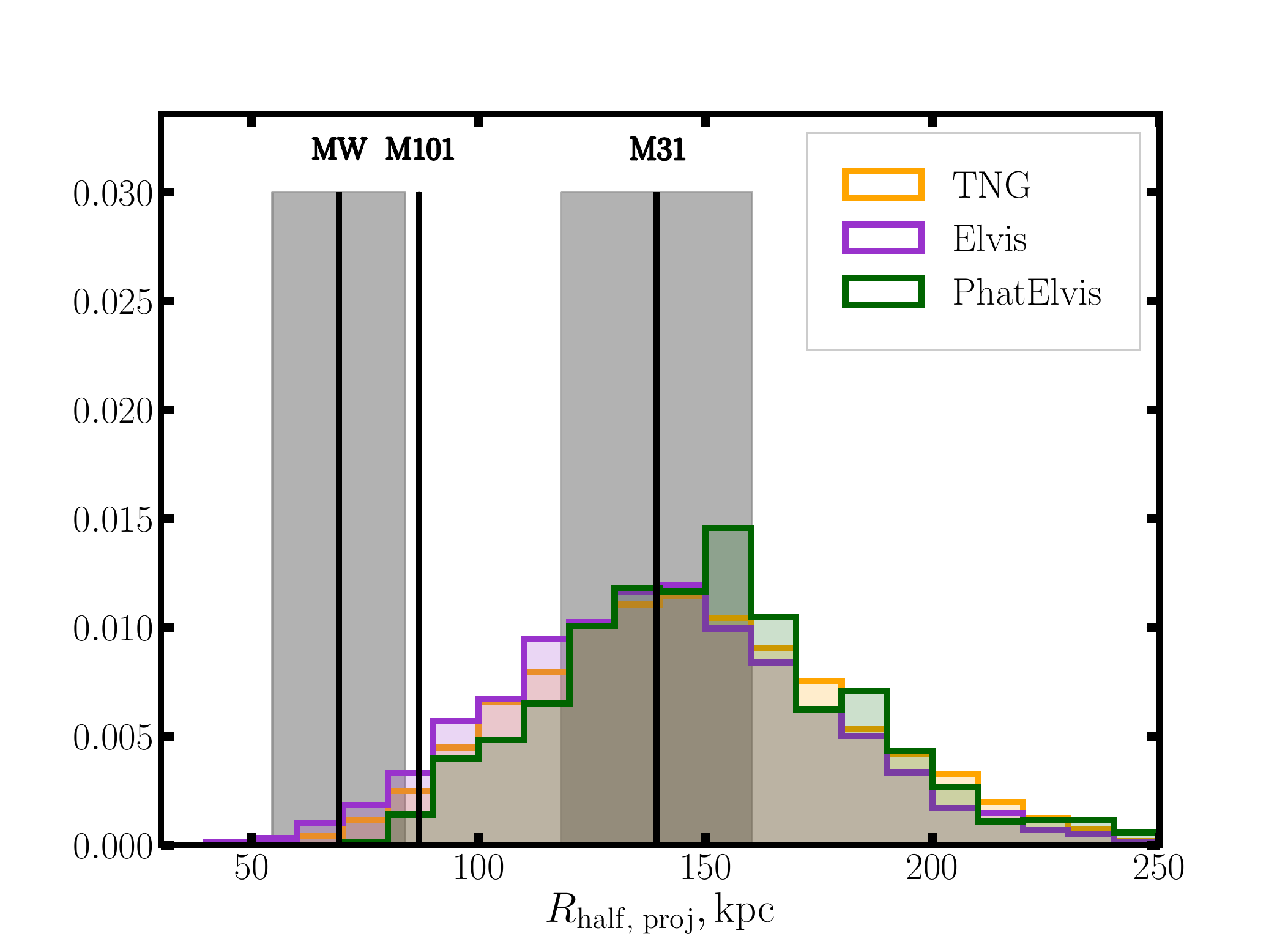}
\caption{The same visualization as in the right panel of Figure \ref{fig:r50} but for satellites within 300 projected kpc. Only the MW, M31, and M101 are shown as these are the only observed systems complete at these radii. The shaded bands show the $\pm1\sigma$ spread in $R_\mathrm{half,\;proj}$ due to assuming different viewing angles for the MW and M31. }
\label{fig:r50_300}
\end{figure}

In Figure \ref{fig:r50_300}, we show the distribution of $R_\mathrm{half,\;proj}$ for the simulations and observed systems when including satellites to 300 projected kpc ($\sim$ the virial radius). Only the MW, M31, and M101 are included as no other observed system is complete to these radii. We include satellites with $M_V<-8.5$ mag which is the estimated completeness for M101. The average number of satellites among these three hosts is 12, and so, for this comparison, we draw the 12 most massive subhalos from the simulations for each host. When including the full system of satellites, the discrepancy between the MW's and M101's systems and the simulated systems becomes much more significant compared to Figure \ref{fig:r50}. This clearly demonstrates the importance of satellite searches that cover the {\it full} virial extent of the hosts.

As a check, we evaluate if our conclusions change if, instead of comparing at a fixed radius, we normalize the satellite separations by the individual virial radii of the hosts. In Figures \ref{fig:indiv}-\ref{fig:r50}, we have considered the satellite separations in physical distances. This is justified by the fact that the hosts we consider all have similar expected virial radii, particularly so when we split the hosts into the MW-like and small-group categories.  Still, it is useful to investigate if the increased satellite concentration relative to the simulated hosts is due to some observed hosts having small virial radii (and, hence, naturally more compact satellite configurations). We found that the results are qualitatively the same when scaling by the host virial radius.
%In Figure \ref{fig:r50_scaled}, we show the distribution of median satellite separations of the simulated and observed hosts, now scaled by the virial radii of the hosts. See \citet{LV_cat} for an estimate of the virial radii of the observed hosts. Only satellites within $r_{\mathrm{proj}}/R_{\mathrm{vir}}<0.5$ are included. This is analogous to the $r_{\mathrm{proj}}<150$ kpc cutoff used in Figure \ref{fig:r50} as 300 kpc is the characteristic virial radii of hosts this size. We assume that the included hosts are complete within $r_{\mathrm{proj}}/R_{\mathrm{vir}}<0.5$. Since M104 is quite a bit more massive that the other observed hosts with a substantially larger estimated virial radius of $\sim600$ kpc, we do not include it. Overall, we see a very similar result as shown in Figure \ref{fig:r50}. The simulations are somewhat less consistent with each other, but that is possibly due to the different definitions of virial radii used in the different suites. $R_{200}$ is used for the IllustrisTNG hosts whereas both ELVIS and PhatELVIS use a \citet{bryan1998} definition ($R_{\sim 100}$). 

\subsection{Comparison to SAGA Survey}
\label{sec:saga}

\begin{figure}
\includegraphics[width=0.5\textwidth]{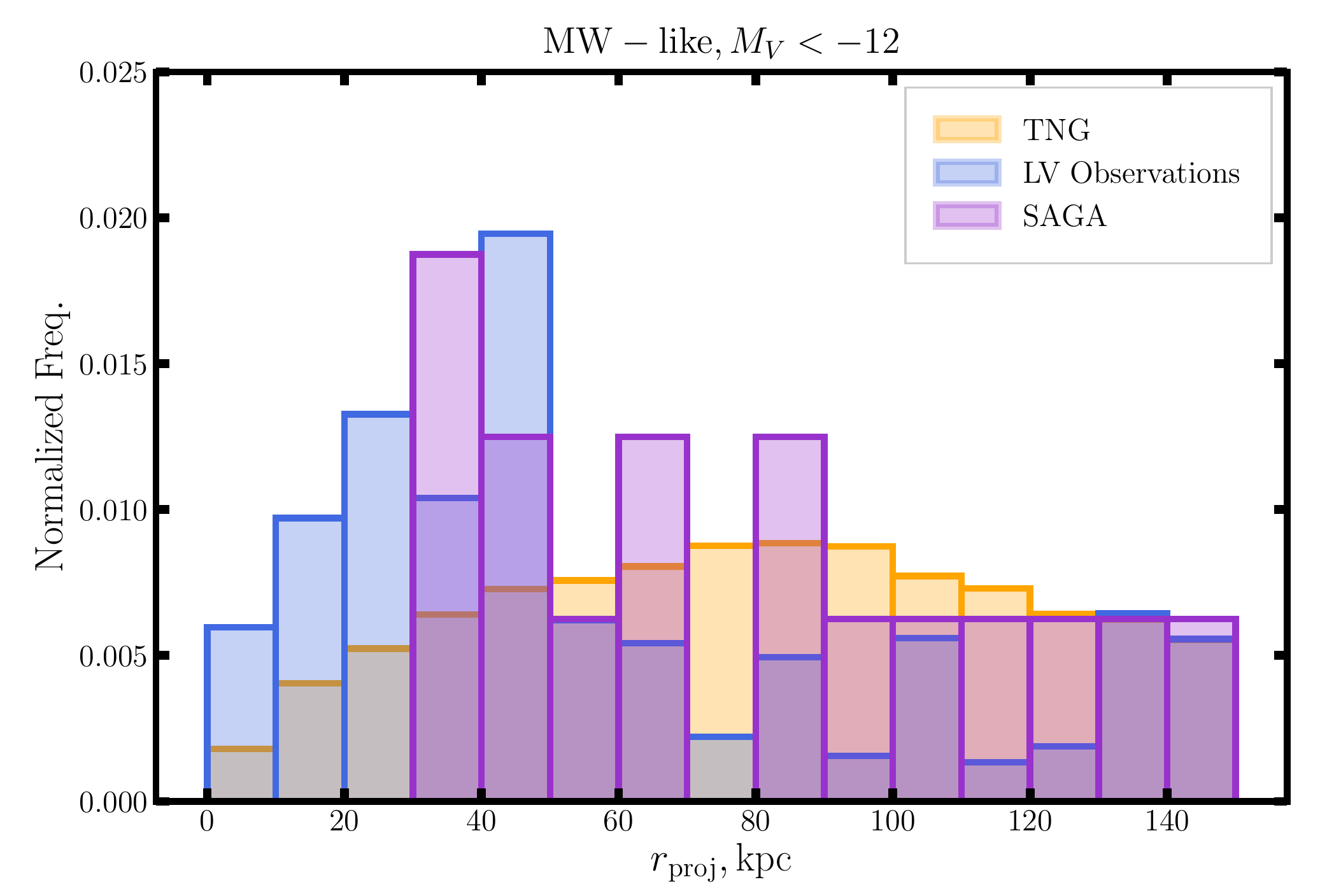}
\caption{The distribution of projected separations for the LV observed satellite systems compared with both the simulated hosts of the IllustrisTNG project and the SAGA Survey. The normalization of the histograms is arbitrary. The IllustrisTNG hosts are processed through the observational selection functions of the LV hosts using the survey area footprints of those hosts. }
\label{fig:rproj_saga}
\end{figure}

In this section, we compare our results to the radial profiles inferred from the SAGA Survey \citep{geha2017}. SAGA focuses on MW-analogs over the distance range $20-40$ Mpc. As such, SAGA is only sensitive to the brightest satellites ($M_V\lesssim-12$); it will, however, in the end have many more observed hosts than what is possible in the LV. \citet{geha2017} presented results for the first eight completed hosts. While the statistical sample of satellites is larger using our LV systems (particularly because our LV hosts are complete to much fainter satellites), it is interesting to compare the SAGA radial profiles with those inferred in this work. Figure \ref{fig:rproj_saga} shows the distribution of projected separations for the MW-like LV hosts for bright ($M_V<-12$) satellites compared both to the confirmed SAGA satellites and to the profiles predicted from the IllustrisTNG hosts; the latter is selected by halo mass in the range $0.8\times10^{12}<M_{200}<3\times10^{12}$ \msun. To maintain our focus on the shape of the distributions, all histograms are normalized. As before, the IllustrisTNG hosts are passed through the observational selection functions of the LV hosts.

Interestingly, the radial distribution of the SAGA satellites is intermediate to the LV host satellites and the IllustrisTNG results. The SAGA surveys does show the same large excess of satellites at $r_\mathrm{proj}\sim40$ kpc, however, there are no very satellites within $R_{proj}$ \textless\ 30 kpc in the current SAGA sample, possibly due to the SDSS-based targeting strategy and the distance of these hosts. It is possible that these very close-in satellites are confused with the outskirts of the hosts, and at the large distance of these hosts, it is hard to distinguish them. We note that the statistical sample of the SAGA satellites is still small and a firm conclusion can not be drawn; in particular, the highest peak at $r_\mathrm{proj}\sim40$ kpc corresponds to only three SAGA satellites. Future SAGA results will provide confirmation of this trend.

\subsection{Comparison To Recent Work}
\label{sec:recent_work}

Recently, \citet{samuel2020} extensively compared the satellite radial profiles of the MW and M31 to simulated hosts in the FIRE project. \citeauthor{samuel2020} argued that the simulated hosts had radial profiles that were consistent with those of the MW and M31. Although we draw different conclusions here, we would argue that the results are fairly similar. First, in our work, we have a statistical sample of hosts. If we only compared the MW and M31 to a handful of simulated hosts, the disagreement would be less egregious because M31's radial distribution is consistent with the simulations. Second, we identify the disagreement {\it only} when we normalize the radial profiles specifically to focus on their shape. In contrast, \citet{samuel2020} primarily consider the absolute radial profile, and the spread of the absolute radial profiles from the models does, in fact, encompass the MW observations. Particularly, this occurs because absolute radial profiles confuse the richness of a satellite system with its spatial distribution. Precisely because we do not use an SHMR to populate subhalos  (instead, we take the same number of massive subhalos as the observed number of satellites), the richness of the simulated systems in our comparison exactly matches that of the observed hosts, by definition. But in the FIRE-2 simulations, the number of satellites per simulated MW-like host differs by a factor of a few between hosts. This spread is able to encompass the centrally concentrated profile of the MW; we demonstrate this explicitly in Appendix \ref{app:norm}. 

\citet{samuel2020} explored the shape of the radial profile of MW and M31 satellites with the ratio of the radius containing 90\% of satellites to that containing 50\%. With this metric, they found that the MW was more concentrated than all of their baryonic simulations, consistent with what we find. In this paper, we opt to use the simpler $R_\mathrm{half}$ as a measure of the concentration of a profile. Because we have systems with only a few satellites, it becomes difficult to define the radius that contains 90\% of satellites.

Our census of MW satellites is different from \citet{samuel2020}. We include Sgr while they do not. \citeauthor{samuel2020} do include Crater II and Antlia II while we do not include these satellites due to their extremely low surface brightness levels that fall below our completeness estimate for the LV host sample that we adopt \citep{LV_cat,lv_lfs}. These small differences have the effect of somewhat boosting the central concentration of MW satellites in our results. Were we to adopt the \citeauthor{samuel2020} sample, however, our conclusions about the relative concentration of the ensemble of hosts would not change significantly because this change only applies to the MW.

Finally, \citet{samuel2020} found that if the the simulated profiles were normalized to that of the MW at $r=150$ kpc, all of the simulated hosts had more satellites at $r>150$ kpc than the MW. They use this to argue that there could be classical-sized satellites yet undiscovered in the periphery of the MW virial volume. We discuss the possibility of this below in \S\ref{sec:disc_obs}.

\subsection{Summary}
\label{sec:comp_summary}
In this section, we performed comparisons between the radial distributions of observed satellites in the LV and analogous simulated systems in a set of modern cosmological simulations. While some of the observed hosts (particularly the more massive `small-group' hosts) seem to have similar radial profiles as the simulated hosts, on average the observed hosts are more concentrated. In Figures \ref{fig:indiv} and \ref{fig:r3d}, we demonstrated this discrepancy by directly comparing the normalized radial distributions of satellites in the observed and simulated hosts. Normalizing the profiles is essential to see the difference between observations and simulations. In Figure \ref{fig:rproj}, we compared the distribution of satellite projected separation, $r_\mathrm{proj}$, between the observed hosts and the simulations. For the MW-like hosts, the distribution of observed satellites is significantly more centrally concentrated than the simulated hosts. This is particularly striking (i) compared to the PhatELVIS simulations and (ii) when only the bright ($M_V<-12$) satellites are considered. Although, we note that the discrepancy is clear for all simulation suites and considering all satellite luminosities. Next, we used the half-satellite radius (the radius enclosing half of the satellites), $R_\mathrm{half}$, to quantify the concentration in the radial profiles. Figure \ref{fig:r50} shows that the population of observed hosts is shifted to smaller values of $R_\mathrm{half}$ than the simulated hosts. While three of the observed hosts show the average value for the simulations, the rest of the observed hosts are more concentrated, and none are less so.

\section{Discussion}
\label{sec:disc}
In this section, we explore possible causes of this discrepancy in the $\Lambda$CDM paradigm. We leave an exploration of non-vanilla cosmological explanations to future work. We start by discussing the observations then move to the simulations.

\subsection{Observational Incompleteness}
\label{sec:disc_obs}
The most obvious possible problem with the observations is that of incompleteness. Here, we describe some of the possible concerns and argue that observational completeness is not the cause of the discrepancy.

The most significant worry is that the LV satellite surveys are less complete at large radii than they are closer to the host. This could be due, for instance, to the dithering pattern of the observations of the central host. Indeed several of the hosts of \citet{LV_cat} had dithering patterns in the CFHT data used there that led to somewhat deeper exposures near the host than at larger $R_{proj}$. However, there are three lines of reasoning indicating that this is not significantly affecting the observed radial profiles. 

First, Figure \ref{fig:rproj} shows that the discrepancy does not get better (in fact it gets worse) when we only consider the brightest satellites that are more likely to be complete. 

Second, the completeness estimates from \citet{LV_cat} (see their Figure 3) indicate that the satellite catalogs drop from $\sim100$\% completeness to $\sim0$\% over roughly half a magnitude for both surface brightness and total magnitude of the satellite. These estimates average over the entire survey footprint and, if there was a significant difference in depth between different areas of the survey, then we would expect the completeness to drop off more gradually. From \citet{LV_cat}, the completeness limit only changed by $\sim1$ mag (but always $M_V\gtrsim-9$) between hosts that had exposure times differing by up to a factor of two, also suggestive that dithering would not lead to a significantly different completeness in the inner and outer regions. 

Finally, we have used DECaLS \citep{decals} to search for satellites in several of the systems included in the present work. The details of these searches will be presented elsewhere (see Appendix \ref{app:extra_data} for a description of our search of NGC 4631). DECaLS is an extremely wide-field imaging survey with uniform depth, and we can easily detect satellites down to $\sim\mu_{0,V}=26$ mag arcsec$^{-2}$, roughly the limit of the current observational sample. With DECaLS we recover the satellites discovered in the CFHT data of \citet{LV_cat}, but do not find any extra satellites further out from the hosts, indicating that non-uniform depth in the CFHT data is not causing a radial bias in the satellite census.

Due to our position within the MW, the MW's satellite census is particularly vulnerable to incompleteness at large radii. Previous authors \citep[e.g][]{yniguez2014,samuel2020} have suggested that the concentration of MW satellites is indicative of incompleteness in the satellite census at large radii. Such missing classical-sized satellites, however, would all have to be in the zone of avoidance around the MW disk to have evaded discovery thus far. The rest of the sky has been searched at a depth that would easily discover classical satellites throughout the virial volume \citep[e.g.][]{whiting2007,koposov2008, tollerud2008, walsh2009, drlica2019}. Obscuration by the disk should not preferentially hide distant satellites, and so it is unclear if this would actually decrease the concentration of the MW satellites. Not finding disk-obscured satellites would, however, increase the scatter of the simulated profiles, perhaps making the MW less of an outlier. Testing this, we find that Figure \ref{fig:r3d} hardly changes if we select only subhalos that are $>15\deg$ above or below a randomly oriented disk in the simulated hosts (as viewed from the center of the host). Even excluding potentially obscured satellites, the MW remains a significant outlier from the simulations.

Another possible issue with the observations is the confirmation of satellites with SBF distances. While our SBF methodology and calibration have been verified with independent \emph{HST} distances \citep{sbf_m101, bennet2019}, the distances are much less secure than \emph{HST} TRGB distances. It is possible that a few unrelated background interlopers are included in the confirmed satellites of \citet{lv_lfs}. However, including background galaxies will make the radial profiles \emph{less} centrally concentrated not more. Additionally, we account for the possibility of dwarfs that are nearby ($\lesssim1$ Mpc) to the host but actually outside of the virial radius by using a generous cut along the line of sight when selecting subhalos. SBF distances are not precise enough to distinguish these objects from bona-fide satellites within the virial radius of the host. From our experience with the simulations, we do not expect these to be a significant source of contamination ($\lesssim10\%$ of the total satellite sample) \citep[see also][]{lv_lfs}. 

When surveying nearby disk systems, especially face-on disk galaxies, it is possible that the surveys miss very close-in satellites because they project onto the face of the host galaxy. However, including these satellites would make the observed profiles \textit{even more} centrally concentrated, not less. In other words, the concentration of the observed systems is a lower bound taking this into account.

\subsection{Reionization}
\label{sec:reion}
Many previous works have argued that the \textit{luminous} subhalos are more centrally concentrated than the DM subhalo population as a whole. The physical basis of this argument is that reionization suppresses star formation in many halos, and the halos that form earliest have the highest chance of being luminous \citep[we note that there are other criteria one could use to decide which subhalos are luminous, e.g.][]{hargis2014}. These subhalos are more concentrated around the host. As mentioned in the Introduction, classical-sized satellites which are the focus of this paper should be above the scale at which reionization can keep a subhalo dark and this is unlikely to affect the radial distribution of classical satellites. 

\citet{font2011} and \citet{starkenburg2013} presented semi-analytic models (SAM) of galaxy formation coupled with DMO simulations and found that the radial distribution of model satellites was similarly concentrated as those of the MW. However, \citet{font2011} considered the radial distribution of UFDs along with the classical satellites. The spatial distribution of UFDs might certainly be biased by reionization, and that is causing the model satellite distribution to agree well with observations. The SAM of \citet{starkenburg2013} implemented the effect of reionization using the filtering mass approach of \citet{gnedin2000} which is now known to over-predict the suppressing effect of reionization \citep[e.g.][]{okamoto2008}. The good agreement that \citet{starkenburg2013} find between their model classical satellites and the observed classical satellites possibly is a result of the stronger suppressing effect of reionization in their implementation, causing a radial bias in the locations of luminous satellites even at the classical mass scale. Modern hydrodynamic simulations seem to support the idea that the suppressing effect of reionization should not be important at the mass scales of the classical satellites \citep{okvirk2016,sawala2016b,wheeler2019}. The SHMR of \citet{gk_2017} indicates that satellites of the luminosity we consider ($M_V\sim-10$) are hosted by halos of mass $\gtrsim5\times10^9$\msun\ which are above the effects of reionization.

Furthermore, the hydrodynamic simulations, which will include the effects of reionization, of the NIHAO project show similar radial distributions to the other DMO simulation suites we compare with, indicating that reionization is not particularly important at the satellite mass scale we deal with here.

\subsection{Effect of Host Assembly History}
\label{sec:host_prop}
\begin{figure}
\includegraphics[width=0.5\textwidth]{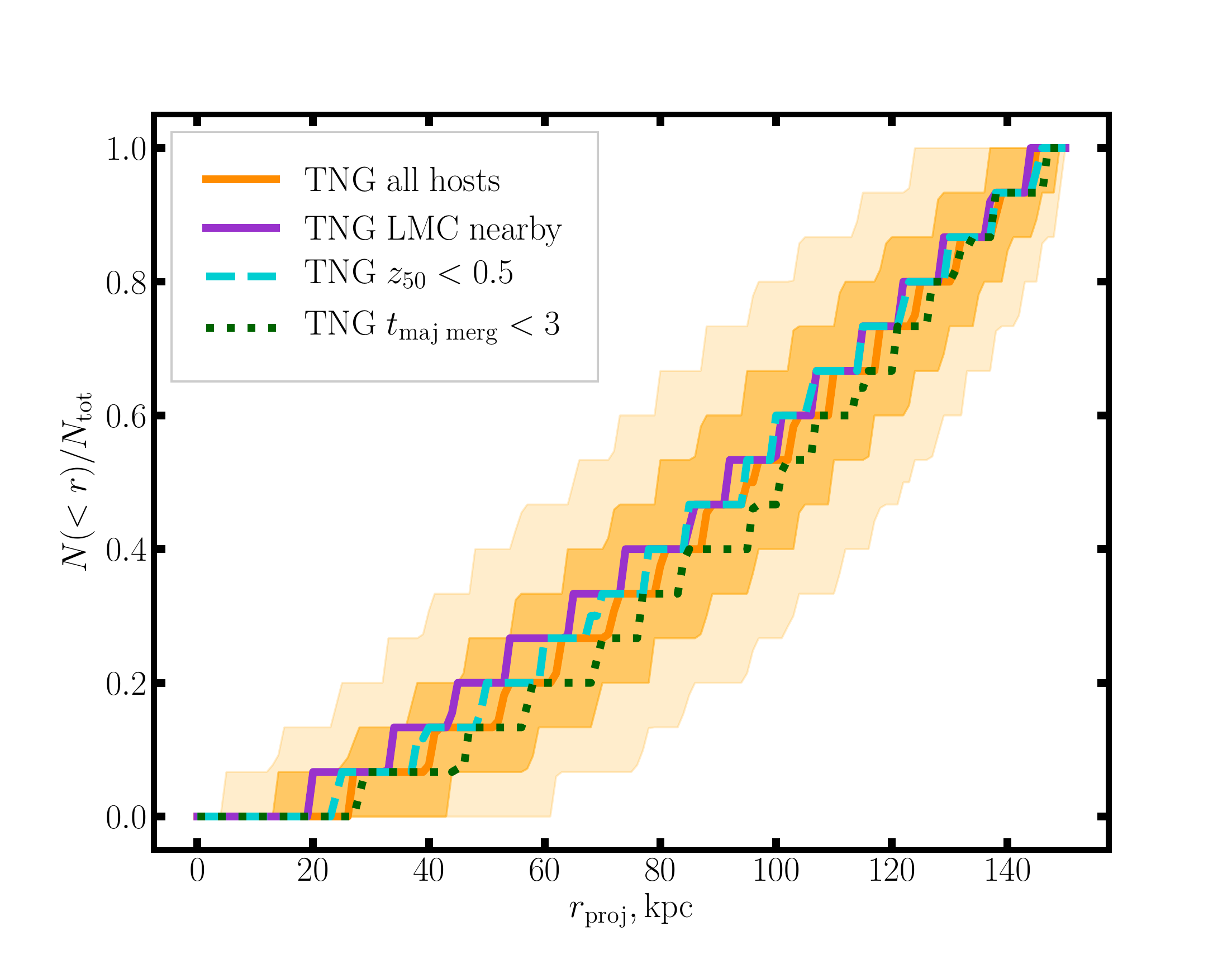}
\caption{The radial distribution of the 15 most massive subhalos in IllustrisTNG hosts within 150 kpc. The orange shows the overall sample of `MW-like' hosts shown in the previous figures. The purple shows the radial distribution for hosts with a Magellanic Cloud-like satellite. The dashed turquoise shows the distribution for hosts that reached half of their current mass after $z=0.5$. The dotted green shows the median profile for hosts that experienced a major merger in the last 3 Gyr. }
\label{fig:rproj_lmc}
\end{figure}

In this section, we explore whether different properties of the simulated hosts have an effect on their radial distribution of subhalos. It is possible that the relatively small sample of observed hosts are biased in some way (e.g. are located in early forming halos). There is significant literature discussing the impact of having a subhalo as massive as that hosting the LMC \citep[e.g.][]{lu2016, nadler2019b} on the overall population of subhalos. So for one check, we select hosts from IllustrisTNG that have a Magellanic Cloud-like satellite. To do this, we select the subsample of `MW-like' halos that have a massive subhalo with $v_\mathrm{max}>55$ km/s \citep[following][]{lu2016} within the distance range $30<D<60$ kpc from the host. This corresponds to roughly 10\% of all halos in the mass range $0.8\times10^{12}<M_{200}<3\times10^{12}$ \msun. Figure \ref{fig:rproj_lmc} shows the radial profiles of these hosts compared to the overall sample of MW-like halos. The profiles are more centrally concentrated for hosts with a MC-like satellite but not nearly at the level of concentration required to explain the MW and other LV hosts. In \citet{lv_lfs}, we found that the LV hosts did show a preponderance of bright (MC-like) satellites compared to the simulated systems using a stellar-to-halo mass relation to populate subhalos. It is unclear whether this is related to the overly concentrated satellite systems we find in this work. We also show the radial profiles for hosts that reached half of their current mass after $z=0.5$ and hosts that experienced a major merger (mass ratio 3:1 or less) in the last 3 Gyr. These criteria were chosen to be about as selective as the presence of the LMC (roughly 10\% of all halos). Neither of these selection criteria lead to a large difference in radial distribution of subhalos. 

We emphasize again here that the current set of observed hosts is $\sim80$\% volume complete for massive ($M_\star\gtrsim1/2\times M_\star^{\mathrm{MW}}$) hosts with $|b|>15^\circ$ within $\sim8.5$ Mpc so we believe it is unlikely that a bias due to a specific host formation history (or any other kind of bias) could be causing the discrepancy. The entire $D\lesssim8.5$ Mpc volume would seemingly have to be an outlier which seems unlikely. However, see \citet{neuzil2020} for possible evidence that this volume is indeed a $\sim2-3\sigma$ outlier compared to simulations with its relatively high density of bright ($M_B<-20.5$) galaxies. With this said, 12 hosts is still a relatively small sample that does not fully span the possible range of environments and host properties, and an increased observational sample is urgently needed.

\subsection{Incompleteness in the Simulations}
\label{sec:disc_sim}
The most likely possible problems with the simulations stem from resolution effects. On the surface, the fact that we get similar answers across the different simulations suites (once we recall that only some of the suites include a central disk), would seem to indicate that resolution is not playing a major role. The suites differ by roughly a factor of 200 in resolution between IllustrisTNG and PhatELVIS. However, it is possible that even if the simulation results are converged with respect to resolution, they might not be converged to the correct answer. We focus on two specific possible resolution-related problems in the application of the simulations.

The first possible problem in the application of the simulations is that we do not allow for the possibility of `orphan galaxies'. Orphan galaxies represent the possibility that the DM halo associated with a luminous satellite becomes stripped to the point that it falls below the threshold of the subhalo finder. The luminous galaxy needs to be manually put back into the simulation and tracked as it has no corresponding DM subhalo. Previous works \citep[e.g.][]{gao2004, newton2018, bose2019} have found that reproducing the radial distribution of luminous satellites in clusters and around the MW required inserting and tracking orphan galaxies. Different prescriptions are used to track the evolution of the orphan galaxies, but generally they are removed after a certain time to represent their merging into the central primary. 

It is unclear whether inserting orphan galaxies is an appropriate thing to do for the comparisons we do in this work. Orphan galaxies are mainly important for subhalos near the resolution limit of the simulation, which is not the case for this work where we are focused on the fairly massive classical-sized satellites. A typical classical satellite hosting subhalo with mass $\sim5\times10^9$ \msun will be resolved with $>10^5$ particles in the PhatELVIS simulations. For this subhalo to drop below the threshold of a subhalo finder ($\sim20$ particles), it has to be $>$99.9\% stripped. By this point a \emph{significant} fraction of the stars would also be stripped \citep[e.g.][]{penarrubia2008} and the satellite would likely either be undetectable (due to very low surface brightness) or clearly tidally disturbed. Some of the observed satellites that we include are clearly undergoing tidal disruption (e.g. Sgr, NGC 5195, NGC 4627, M32, NGC 205, dw1240p3237) but the majority are not. Not including these satellites will clearly reduce the observed central concentration but not enough to make the discrepancy go away. 

On the other hand, it is not clear that these disrupting systems would not be captured by the halo finders in the simulations. Sgr is estimated to still have a significant DM halo from the velocity dispersion of stars \citep{law2010} that would be easily resolved in the simulation suites we use. However, perhaps there is not enough contrast between its DM subhalo and the parent halo for the subhalo to be identified as a distinct structure by a halo finder. NGC 205 similarly is estimated to have a significant component of dark mass \citep{geha2006}. Therefore, it does not seem legitimate to include a large population of orphan galaxies as these heavily stripped galaxies do not correspond to the population of observed satellites. More work is required to fully understand the correspondence between observations and simulations for disrupting satellites at various stages of disruption to understand when such systems will no longer be observable or detected by halo finders in the simulations. With this said, we explore a model with a population of orphan galaxies in the next section.

The second, related possible problem with the simulations is the possibility of a significant fraction of the tidal disruption of subhalos in the simulations being artificial. \citet{vdb2018a} and \citet{vdb2018b} argue that even for cosmological simulations that are `converged' and resolve subhalos with $>100$ particles, most of the tidal disruption of these subhalos is artificial. They cite discreteness noise and inadequate force softening as the main culprits. \citet{vdb2018b} provides new criteria to judge whether the tidal evolution of a subhalo is trustworthy or whether it is affected by discreteness noise. For a subhalo on a circular orbit with a radius of 0.2 times the host virial radius, the subhalo needs to be resolved with $>10^5$ particles. For the high-resolution PhatELVIS simulations, the subhalos that host bright classical-sized satellites would be resolved at roughly this level. However, the \citet{vdb2018b} criteria are for circular orbits in a NFW tidal field, and it is unclear how to extend this to non-circular orbits in the presence of a disk's tidal field. Artificial disruption of subhalos could easily explain the discrepancy we find if many of the simulated subhalos near the host galaxy were artificially destroyed.

If the halo catalogs are missing objects, either orphans or artificially disrupted halos, then the total number of predicted halos would necessarily rise, leading to the need for a new SHMR. In \citet{lv_lfs}, we found that the observed LV satellite systems have total abundances well matched by the simulations combined with standard abundance matching relations found in the literature, in particular, that of \citet{gk_2017}. The SHMR of \citet{gk_2017} shows fair agreement with modern hydrodynamic simulations. However, if a significant number of artificially lost subhalos need to be added into the simulations, a much steeper slope or lower normalization of the SHMR will be required to not over produce satellites in the simulations. We show this qualitatively with a toy model for orphan galaxies in the next section.

\subsection{Orphan Galaxies Toy Model}
\label{app:orphan}
While it is unclear how to resolve the discrepancy found in this paper, as discussed in the last section, a possible option is that the simulations are missing subhalos, either from artificial over-merging or incompleteness in the halofinder perhaps due to resolution effects. If the simulation halo catalogs are missing a large population of close-in subhalos, putting these back in could help resolve the discrepancy. However, adding more subhalos would effect what stellar halo mass relation (SHMR) is allowed by the observed abundances of satellites. In \citet{lv_lfs}, we showed that the overall richness of the LV satellite systems could be well matched by the SHMR of \citet{gk_2017}, however, the model predicted too few bright satellites and too many faint satellites. This SHMR agrees well with the predictions from recent hydrodynamic simulations of dwarf galaxy formation. If a large population of subhalos is added into the simulations, this SHMR likely will no longer match the observations. In this section, we develop a toy model that inserts subhalos into the halo catalogs to match the observed radial distribution, and we investigate the effects on the SHMR allowed by observations.

\begin{figure*}
\includegraphics[width=1.02\textwidth]{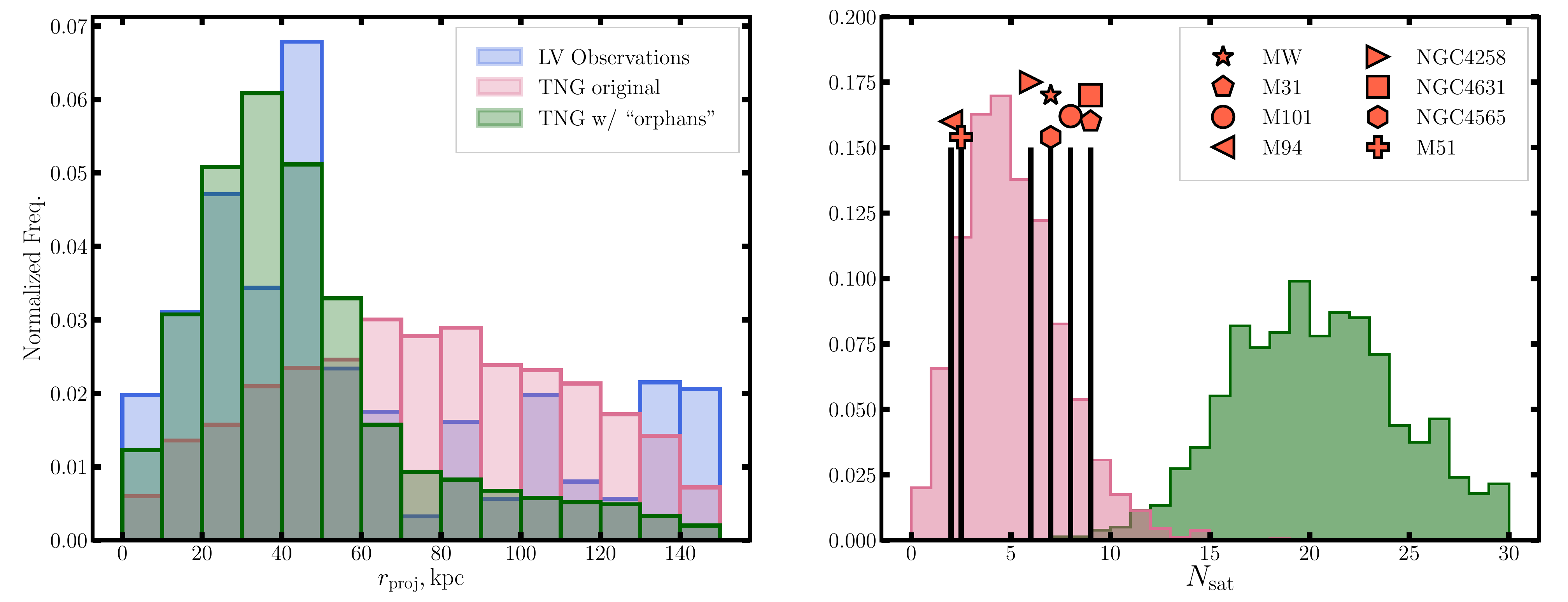}
\caption{One possibility to resolve the discrepancy in radial distributions between simulated and observed hosts is that the simulations are missing many subhalos, either due to artificial disruption or halofinder incompleteness. This figure shows the effect of adding these subhalos back into the simulation. \textit{Left:} The radial distribution of the IllustrisTNG simulations after these `orphan' galaxies are added back in. See text for details about how they are added. \textit{Right:} The predicted abundance of $M_V<-9$ satellites within $r_\mathrm{proj}<150$ kpc using the SHMR of \citet{gk_2017} with and without the added subhalos. As expected, injecting subhalos would require a significantly steeper slope or lower normalization for the SHMR to avoid over-producing satellites.}
\label{fig:shmr}
\end{figure*}

For this test, we use the TNG simulations. We only consider the 8 MW-sized observed hosts in this comparison and compare with TNG hosts selected by halo mass in the range $0.8\times10^{12}<M_{200}<3\times10^{12}$\msun. While other treatments of orphan galaxies insert them and trace their movement in the host halo in a physically motivated way, we insert the orphan galaxies directly in the halo catalogs. We insert the orphans with a projected separation from the host drawn from a Gaussian with mean 35 kpc and standard deviation of 15 kpc. These numbers are chosen to match the peak of close-in observed bright satellites visible in Figure \ref{fig:rproj}. This is clearly unphysical. This is just a toy model which we are using to answer the question: ``would including the number of orphan galaxies required to bring the observed and simulated radial distributions into agreement affect the SHMR?" We leave an investigation of where the orphan galaxies would actually end up in the simulations to future work. Subhalo masses are drawn from a distribution between $10^{8.5}<M/M_\odot<10^{11.8}$ using the well-known subhalo mass function $dN/dM\propto M^{-1.9}$ \citep[e.g.][]{springel2008}. Using this distribution assumes that subhalos are artificially lost equally at all masses. This is likely a lower bound as lower-mass subhalos are most likely lost preferentially if this is due to resolution\footnote{On the other hand, the more massive satellites will experience more dynamical friction, and may naturally be more concentrated than the whole satellite population.}. We manually adjust the number of subhalos that we add in to match the distribution of bright satellites (top right corner of Figure \ref{fig:rproj}). We find that adding $\sim100$ subhalos in this mass range within the inner 150 projected kpc leads to fair agreement with the observations. Figure \ref{fig:shmr} shows this result on the left. The observed radial distribution of bright ($M_V<-12$) satellites are shown along with the distribution of the 3 most massive subhalos before and after subhalo injection. Recall that there are, on average, 3 satellites in this luminosity range per observed host, and so we select the 3 most massive subhalos as the simulated satellites. Note that we do not do any survey area corrections for the simulations in this test; we assume the observed systems are all complete to a projected 150 kpc.

On the right of Figure \ref{fig:shmr}, we show the resulting predicted abundance of $M_V<-9$ satellites within $r_\mathrm{proj}<150$ kpc using the halo catalogs with and without injected subhalos and the SHMR of \citet{gk_2017}. Without any added subhalos, we see the result found by \citet{lv_lfs} that the SHMR of \citet{gk_2017} produces roughly the right number of satellites (including host-to-host scatter). However, with the added subhalos, the SHMR significantly overproduces satellites. A significantly steeper slope or lower normalization for the SHMR would be required. It is possible that the required SHMR would no longer be in agreement with the results of modern hydrodynamic simulations of dwarf galaxy formation, but we leave a detailed investigation of this to future work.

\section{Conclusions}
\label{sec:concl}
In this paper, we compare the observed satellite spatial distributions in 12 LV massive hosts to that predicted in various state-of-the-art cosmological simulations in the $\Lambda$CDM paradigm. While many previous works compared the radial distribution of satellites around the MW and/or M31 to simulated analogs, we are able to achieve much better statistics by considering the satellite systems of many MW-like galaxies in the LV. This has only recently been made possible by observational work characterizing the satellite systems of nearby MW analogs. We compare the observations to multiple different simulation suites. These include a big-box cosmological simulation (IllustrisTNG-100) that gives great statistics with $>1000$ MW-like hosts but at relatively low resolution, high resolution DMO zoom-in simulations of several tens of MW-sized hosts both including the potential of a disk (PhatELVIS) and not (ELVIS), and a high resolution fully hydrodynamic zoom in simulation of 6 MW-like hosts (NIHAO). Overall we find fairly good agreement amongst the simulations. Our main findings are as follows:
\begin{enumerate}[label=(\roman*)]
\item We confirm previous findings that the classical satellites of the MW are significantly more concentrated at a $>2\sigma$ level than the massive subhalos of simulated analogs (Figure \ref{fig:r3d}). We argue that this discrepancy is likely not resolved by either reionization or incompleteness in the census of MW classical-sized satellites. 
\item We find that several of the other observed hosts in the LV have more concentrated radial profiles than the analogous simulated hosts at the $\sim2\sigma$ level (Figure \ref{fig:indiv}). A few of the observed hosts, particularly the more massive (`small-group') hosts such as M81 and CenA, have similar radial profiles to the simulated hosts, but \emph{none} of the observed are less concentrated than the average simulated profile. 
\item We use the median satellite separation for satellites within a projected 150 kpc as a metric for the concentration of the satellite spatial distribution. We find that the population of observed systems is systemically shifted to smaller radii compared to the simulated systems (Figure \ref{fig:r50}).
\item Figure \ref{fig:rproj} shows that the spatial distribution of the satellites is significantly different between the observed and simulated satellites. There is a significant population of close in observed satellites that is missing in the simulations. This is particularly noticeable for the bright satellites ($M_V<-12$) where there are a large number of observed satellites at projected separations of $30-60$ kpc but very few simulated satellites at these separations.
\item The spatial distribution of satellites found in the SAGA Survey \citep{geha2017} agrees with the distribution of observed LV satellites, although the statistics are too low to be conclusive (Figure \ref{fig:rproj_saga}). 
\end{enumerate}
Throughout this work we have been careful to consider the incompleteness and limitations of the observations. We have used the specific survey footprints for each observed host, where possible, to forward model the simulated hosts. 

In \S\ref{sec:disc}, we discussed many possible causes for the discrepancy both on the observational side and on the simulation side. There does not seem to be any likely causes for this on the observational side, due to incompleteness. Observational completeness has been well quantified by the various works that have characterized the LV satellite systems.  We do note, however, that many of the discrepant observed systems (e.g. NGC 4631 and NGC 4258) come from the work of \citet{LV_cat}. Excluding these systems makes the discrepancy less severe. With that said, the MW is quite discrepant with the simulations. Since the MW is arguably the observed host with the best completeness, this is suggestive that there is more to this than simple observational incompleteness. On the other hand, the observational sample only contains 12 hosts which is still a relatively small sample and might not fully span the relevant range of environments and formation histories that the cosmological simulations span. 

We also discuss possible causes within the simulations, but no explanation is satisfactory. We discussed whether reionization or biased formation histories of the observed host halos could explain the discrepancy. Reionization has the effect of increasing the central concentration of luminous satellites but is unlikely to be effective at the satellite luminosities considered here. Different halo formation histories (e.g. late major merger, presence of an LMC-like companion at $z=0$, etc.) do have a minor effect on the radial distribution of satellites but not enough to explain the observations. This is an important check as our hosts do have an unusually high prevalence of massive satellites \citep{lv_lfs}.

Finally we discussed possible issues in the simulations. In particular, we do not include the possibility of orphan galaxies in the simulations. These are galaxies whose DM subhalo has dropped below the detection threshold of the subhalo finder and need to be tracked `manually'. We argue that, at the subhalo masses of the satellites we consider, when the subhalo is stripped to the point it is not detected by a subhalo finder, the luminous galaxy would be mostly destroyed and, hence, does not correspond to the observed satellites. We also discuss the possibility of significant artificial disruption in the simulations. This appears to be the most feasible cause of the discrepancy. If this is the cause, this will have important ramifications for the allowable SHMR in this mass range. \citet{lv_lfs} found that the SHMR consistent with state-of-the-art hydrodynamic simulations from various projects \citep[see e.g.][and Figure \ref{fig:shmr}]{gk_2017} reproduces fairly well the overall number of observed satellites in these systems. If there is a large population of subhalos that are getting artificially disrupted in the simulations (but in reality should still exist), then this SHMR will overproduce satellites and the relation will have to be steepened significantly or given a lower normalization.

The observed systems appear to disagree most with the simulated hosts in the PhatELVIS suite. Due to its high resolution and inclusion of a central disk, we expect the PhatELVIS suite to be the simulations that most realistically represent MW-like systems. This highlights the fact that disruption of DM substructure by a central disk is still not entirely understood. Our findings of too-concentrated satellite systems are similar to the results considering the UFDs of the MW where it is difficult to reconcile the abundance of observed close-in UFDs with the dearth of close-in subhalos once disk disruption is accounted for \citep[e.g.][]{kim2018,graus2019,nadler2019b}. We note that the model of  \citet{nadler2019b} was unable to reproduce the radial distribution of UFDs around the MW even when including a population of orphan galaxies. Given the importance of tidal stripping and disruption in the baryonic resolution of both the `Missing Satellites' and `Too Big to Fail' Problems of small-scale structure formation, it is crucial to fully understand the how the central disk affects the population of DM subhalos.

On the observational front, the way forward is still clearly to survey and characterize more satellite systems, and emphasis should be given on surveying a few systems out fully to the virial radius of the host. Much of the significance of the discrepancy of the MW's radial distribution of classical satellites comes from the fact that it is surveyed out to the virial radius (300 kpc). Figure \ref{fig:r50} shows that the inner ($r_\mathrm{proj}<150$ kpc) satellites of the MW are more concentrated than simulated analogs but only at a $\sim1\sigma$ level. In Figure \ref{fig:r50_300}, we show that when considering the entire satellite system out to 300 kpc, the discrepancy between $R_\mathrm{half}$ of the MW and M101 and that of the simulated systems becomes much more significant at $\gtrsim2-3\sigma$. A volume limited sample of hosts in the LV would be a particularly powerful data set to compare with simulations.

Finally, we note that this radial distribution of satellites is not the only observed peculiarity of the spatial distribution of the MW satellites. It has long been known that the classical satellites are arranged in a thin plane \citep{kroupa2005}, the likes of which are quite rare in cosmological simulations \citep[e.g.][]{pawlowskiVPOS,pawlowski2019}. It is certainly possible that the unusual radial distribution of the MW satellites is related to their unusual planar configuration. Other systems (e.g. NGC 4258, NGC 4631, and M101) having similarly concentrated radial profiles makes this possible connection all the more intriguing.

\section*{Acknowledgements}
% -- RLB 
We thank J. Samuel and A. Wetzel for their willingness to share the FIRE-2 simulation results in an earlier version of this work. 

Support for this work was provided by NASA through Hubble Fellowship grant \#51386.01 awarded to R.L.B. by the Space Telescope Science Institute, which is operated by the Association of  Universities for Research in Astronomy, Inc., for NASA, under contract NAS 5-26555. J.P.G. is supported by an NSF Astronomy and Astrophysics Postdoctoral Fellowship under award AST-1801921. J.E.G. is partially supported by the National Science Foundation grant AST-1713828. S.G.C acknowledges support by the National Science Foundation Graduate Research Fellowship Program under Grant No. \#DGE-1656466. AHGP is supported by National Science Foundation Grant Numbers AST-1615838 and AST-1813628.

Based on observations obtained with MegaPrime/MegaCam, a joint project of CFHT and CEA/IRFU, at the Canada-France-Hawaii Telescope (CFHT) which is operated by the National Research Council (NRC) of Canada, the Institut National des Science de l'Univers of the Centre National de la Recherche Scientifique (CNRS) of France, and the University of Hawaii.

\software{ \texttt{astropy} \citep{astropy}}

\bibliographystyle{aasjournal}
\bibliography{calib}

\appendix

\section{New Satellites of NGC 4631 and M101}
\label{app:extra_data}
We have extended the satellite census for NGC 4631 and M101 out to 200 and 300 kpc, respectively, using the extremely wide-field DECaLS imaging \citep{decals}. Details of this satellite search will be given in an upcoming paper (Carlsten et al. in prep), but we give an overview here. We applied the LSB galaxy detection algorithm of \citet{LV_cat} to the $g$ and $r$ imaging of DECaLS to identify candidate satellites. We estimated the completeness of these satellite searches by injecting mock galaxies, following \citet{LV_cat}. We estimate the completeness as $M_V\sim-9$ mag and $\mu_{0,r}\sim26$ mag arcsec$^{-2}$ throughout the search footprint. We then used deeper ground based imaging to measure the surface brightness fluctuations of these galaxies and constrain their distance. The new satellites are listed in Table \ref{tab:new_sats}.

We identified two new candidate satellites around M101 in the 300 kpc radius footprint of our DECaLS search. One of these (dw1403p5338) is actually in the footprint of the search of \citet{bennet2017} but was missed in that search. It was noted by \citet{pvd2019} as a candidate satellite. We used the deep CFHT Legacy Survey imaging to measure its SBF distance which is consistent with M101. We follow the criteria outlined in \citet{lv_lfs} for `confirming' a candidate satellite. This satellite is below the $M_V\sim-9$ fiducial completeness limit that we use for most of this paper and, thus, does not play a large role in the results. The other candidate satellite, dw1350p5441, was outside of the CFHT Legacy Survey footprint but had archival HSC imaging which was deep enough to measure its SBF signal with high S/N. It also is at the distance of M101. 

We identified 6 new satellite candidates around NGC 4631 within 200 kpc but outside of the search footprint of \citet{LV_cat}. We acquired deep ground based images of four of these through the Gemini Fast-Turnaround Queue (Prop ID: GN-2020A-FT-104; PI: S. Carlsten). The imaging was deep enough to constrain three of these to be background contaminants and confirm one (dw1248p3158) to be a real satellite. One of the remaining 2 candidates we could constrain to be background from the actual DECaLS data itself. The remaining candidate was the faintest ($M_V\gtrsim-9.5$) and still is unconstrained. Thus, our completeness with full distance confirmation is roughly $M_V\sim-9.5$ over the inner 200 kpc for NGC 4631.

\begin{deluxetable*}{cccccc}
\tablecaption{Newly confirmed satellites of NGC 4631 and M101. M101 is at $D=6.5$ Mpc \citep{beaton2019} and NGC 4631 is at $D=7.4$ Mpc. \label{tab:new_sats}}
\tablehead{\colhead{Host Name} & \colhead{Dwarf Name} & RA & Dec & $M_V$ & SBF Distance (Mpc)}
\startdata
M101 & dw1403p5338 & 14:03:27.3  &	+53:37:52	&	-8.8  & $6.0\pm0.8$	 \\
M101 & dw1350p5441 &	13:50:58.4 &	+54:41:21	&	-12.0 & $6.4\pm0.8$	 \\
NGC 4631 & dw1248p3158 & 12:48:52.6 & +31:58:13 & -12.5 & $7.1\pm0.8$ \\
\enddata
\end{deluxetable*}

\section{Normalized vs. Absolute Radial Profiles}
\label{app:norm}

\begin{figure*}
\includegraphics[width=\textwidth]{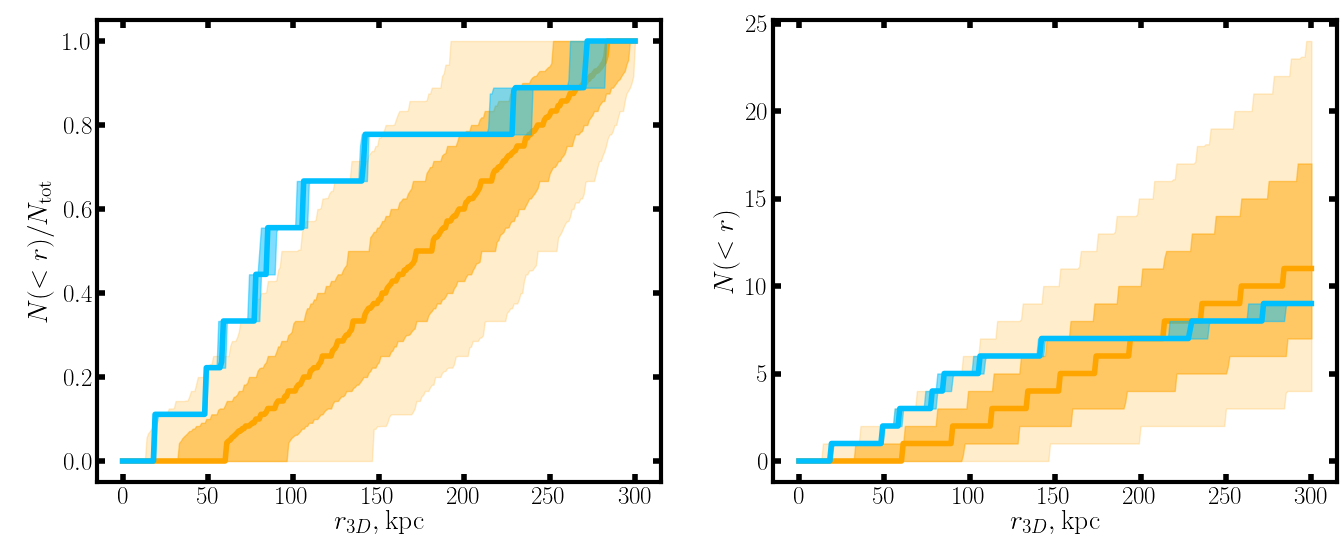}
\caption{The radial distribution of MW satellites ($M_V<-9$) compared to the TNG simulations combined with the SHMR of \citet{gk_2017}.  }
\label{fig:unnorm}
\end{figure*}

Throughout the paper, we consider the normalized radial profiles of satellites instead of the absolute profiles. Since we select the same number of satellites in the simulations as in the observed hosts, the satellite abundance is matched, by definition. However, if we were to use a stellar-to-halo mass relation (SHMR) from abundance matching to populate subhalos with luminous galaxies or used hydrodynamic simulation results, the abundance of satellites would not necessarily be matched, and it would be important to normalize the radial profiles to do a detailed investigation on their shapes. This is part of the reason our conclusions differ from that of \citet{samuel2020} who considered largely the absolute (unnormalized) radial profiles. Absolute radial profiles conflate the problem of satellite abundance and spatial distribution and the scatter in the simulated profiles is much larger and able to encompass (even very concentrated) observed profiles. 

We show this in Fig \ref{fig:unnorm} for the MW satellites ($M_V<-9$). We compare with the TNG simulations combined with the SHMR of \citet{gk_2017}. We assume a $M/L_V$ of 1.2 and only select model satellites brighter than  $M_V<-9$. We show both the normalized and absolute radial profiles. On the right, the simulated satellite abundance differs by a factor of a few between hosts, and this spread is able to encompass the MW's centrally concentrated radial profile.

\end{document}